\documentclass{svjour3}
\usepackage{graphicx}
\begin{document}
\title{\textbf{Towards} $E^{}_{8}\times E_{8}^{\prime}$ \textbf{Unification}}
\author{George Triantaphyllou}
\institute{National Technical University of Athens,\\
             58 SINA Str., GR 106 72 ATHENS, Greece,
              Tel.: +30697 2999816\\
   \email{gtriantaphyllou@aya.yale.edu}}
\date{}

\maketitle

\begin{abstract}
      We present a framework having the potential to unify
  the fundamental interactions in nature  by introducing new
 degrees of freedom.  An attempt is made to explain the 
hierarchy between the weak scale and the coupling unification  scale,
which is found to lie close to Planck energies. A novel process leading
to the emergence of symmetry is proposed, which not only reduces the arbitrariness of the
scenario proposed but is also followed by significant cosmological implications.
Phenomenology  includes the probability of detection 
of mirror fermions via the corresponding composite bosonic states and the relevant quantum corrections at the LHC.
\end{abstract}

\section{Introduction}
\label{sec:1}
Attempts to unify fundamental interactions within a unique theory are not only based on aesthetic and philosophical grounds, but are also reinforced by experiments implying a convergence of the gauge-coupling strengths \cite{Georgi}. We built here upon previous work \cite{Triantap1}\cite{Triantap2} trying to explain fermion family structure and the hierarchy between the electroweak (EW) symmetry scale and the coupling unification scale by introducing new fermions called “katoptrons” (from the Greek word for ``mirror"), since it is imperative to distinguish them from ordinary mirror fermions appearing in alternative models. Stabilization of the EW  scale is achieved  by using a new gauge interaction $SU(3)_{K}$ becoming strong around 1 TeV. The Higgs mechanism is based on fermion condensates, in a spirit close to the study of QCD pions and similar to, but differing significantly from, technicolor \cite{Appelquist}. If katoptrons did not carry  $SU(3)_K$ charges, they would be the mirror partners of the known fermions, the existence of which was first suggested in \cite{Lee} in order to restore the apparent left-right asymmetry in nature. Breaking the electroweak symmetry dynamically by strongly-interacting mirror fermions was previously proposed also in \cite{Wil} by using different gauge symmetries than the ones appearing in the present paper, in order to achieve unification of the gauge coupling strengths, but failed to reproduce the correct weak scale.

Adding so many new particles seems to be required from a theory  involving neither new elementary zero-spin particles, nor new energy scales fixed {\it ad hoc}, nor too many arbitrary parameters. Moreover, this approach can in principle produce testable predictions since the effects of these particles, if they exist at all, are expected to be studied at the LHC \cite{Triantap3}. In order to bring these considerations a step closer to gravity, the initial gauge symmetry was previously enlarged to $E^{}_{8} \times E_{8}^{\prime}$ in $10$ spacetime dimensions, with known fermions contained in $E_8$ and katoptrons in $E_{8}^{\prime}$ \cite{Triantap4}. Since stabilization of extra dimensions is hard, it makes sense to wonder if too many, and hard to detect, degrees of freedom were added in that work. Enlarging the gauge symmetry to such an extent could possibly suffice to bring us closer to gravity, avoiding the addition of extra dimensions by embedding the Lorentz group within a larger symmetry \cite{Percacci}. We proceed here towards this direction by adopting a bottom-up approach, going from lower energies where theories are well tested to higher energies where powerful theorems have to be surmounted, new theoretical methods have to be developed and experimental data are lacking. The constructive-inductive methodology followed should not obscure the ultimate consistency of the emerging picture if one wishes to start from Planck energies and deduce the resulting physical phenomena on the way down to the weak scale.

First, we inquire whether a particular breaking chain of an initial gauge symmetry $G=E^{}_{8} \times E_{8}^{\prime}$ down to the Standard Model is compatible with the unification of the gauge couplings near the Planck scale and leads naturally to dynamical electroweak symmetry breaking at scales of around 1 TeV. We investigate whether the fermion fields surviving at lower energies and the effective composite fields leading to symmetry breaking appear naturally within this setting, a property which is {\it a priori} neither trivial nor obvious. Next, an attempt is made to incorporate the Lorentz group within the initial exceptional-group symmetry $G$ in order to judge whether the model proposed is in principle compatible with models of gravity based on spinors, which were so far based only on orthogonal groups, and which could  potentially allow the inclusion of gravitational interactions in related unification considerations. Then, an effort is made to justify the value of the gauge couplings at the unification scale using a rough order-of-magnitude calculation involving fermion condensates that would enable us to express the hierarchy between the weak scale and the Planck scale in terms of a symmetry-group invariant. Last, a novel mechanism of symmetry emergence is proposed within the framework of critical phenomena and spinor gravity, in order to reduce the arbitrariness of our choice of the initial symmetry $G$. Even though the problems arising in this emergence approach are mostly intractable analytically, one might be able draw useful qualitative conclusions on the critical temperature involved and derive order-of-magnitude results addressing several important cosmological issues like the value of the cosmological constant, the nature of Dark Matter and the interplay between space-time and elementary particles. We proceed below with our first task, {\it i.e.} the analytical calculation, in lowest order, of the relation between the  weak scale and the coupling unification scale.

\section{Coupling unification, symmetry breaking and spinor gravity}
\label{sec:2}
\subsection{Unification and the hierarchy problem}
One of the results of \cite{Triantap1} is coupling unification at high energies, including the $SU(3)_K$ coupling.  The starting point here is a  different initial symmetry $G$ assumed to break at energies $\Lambda_{GUT}$ down to $SU(5) \times U(1)_X \times SU(5)^{\prime} \times U(1)^{\prime}_X \times SU(3)_K$.  We investigate the running of the couplings in this case to see if the energy scales involved have theoretically and phenomenologically acceptable values. 
Under $SU(5) \times U(1)_X \times SU(5)^{\prime} \times U(1)^{\prime}_X \times SU(3)_K$, left-handed ordinary fermions  $F$ and right-handed katoptrons $K$ are taken to transform as follows:
\begin{eqnarray}
F^{a}_{L}  &=& (\bar{\textbf{5}},-3,\textbf{1},0,\textbf{1})^{a} \oplus (\textbf{10},1,\textbf{1},0,\textbf{1})^a \oplus (\textbf{1},5, \textbf{1},0,\textbf{1})^a\nonumber\\
K_{R} &=&  (\textbf{1},0,\bar{\textbf{5}},-3,\textbf{3})^{~} \oplus (\textbf{1},0,\textbf{10},1,\textbf{3})^{~} \oplus (\textbf{1},0,\textbf{1},5, \textbf{3}) 
\end{eqnarray}

\noindent where $a=1,2,3$ is a fermion generation superscript, $\textbf{1}$ and $0$ denote non-abelian and abelian group singlets respectively, ordinary fermions are singlets under $ SU(5)^{\prime} \times U(1)^{\prime}_X \times SU(3)_K$ and appear in 3 generations, while katoptrons are singlets under $SU(5) \times U(1)_X$ and triplets under $SU(3)_{K}$. In an assignment inspired by ``flipped $SU(5)$" models \cite{flipped},  the $\bar{\textbf{5}}$  of  $SU(5)$ contains a lepton doublet and up-type antiquarks, the $\textbf{10}$  a quark doublet, down-type antiquarks and a neutrino, and the $SU(5)$ singlet is a positively-charged lepton (positron etc.), and similarly for the katoptrons in $SU(5)^{\prime}$. 

The symmetry $SU(5) \times U(1)_X  \times SU(5)^{\prime} \times U(1)^{\prime}_X$  is assumed to break at energies $\Lambda_{23}$ down to the Standard Model (SM) group times an abelian $U(1)_{1}^{\prime}$ felt only by katoptrons. One is then left with
$SU(3)_C \times SU(2)_L \times U(1)_1  \times U(1)^{\prime}_1 \times SU(3)_{K}$, where $U(1)_1$ is the hypercharge group with a rescaled coupling. The fields leading to this breaking are neglected in the 1-loop calculation of coupling renormalization below. While katoptrons interact with the same $SU(3)_{C}\times SU(2)_{L}$ interaction as known fermions at energies below $\Lambda_{23}$, they  carry their own $U(1)_{1}^{\prime}$ interaction down to the EW symmetry scale. Moreover,  $SU(3)_K$ becomes strong near  $\Lambda_K$, breaking itself and the EW symmetry \cite{Triantap2}. 

The renormalization of the gauge couplings $g_{N}$ at energy scales $p$ for $N_{f}$ fermions in the fundamental representation (rep) of $SU(N)$ at 1-loop is given by $\alpha^{-1}_N  (p)= \alpha^{-1}_N  (p_o) + c(N, N_{f}) \ln{(p/p_o)}$, with $p_o$ some reference scale, $a_N = g_N^2/4\pi$ and  $c(N, N_f) = (11N-2N_f)/6\pi$. The katoptron coupling $\alpha_{K}$ evolves at scales ranging from $\Lambda_{K}$ to the unification scale  $\Lambda_{GUT}$ according to $c_K \equiv c(3,8) = 17/6\pi$. The   $SU(2)_{L}$ coupling $\alpha_{2}$ and the  $SU(3)_{C}$ coupling $\alpha_{3}$  evolve according to $\tilde{c} _{N} \equiv c(N, 12) = (11N-24)/6\pi$ at energies where both ordinary and katoptron fermions contribute to the beta functions, {\it i.e.} between $\Lambda_{K}$ and $\Lambda_{23}$. Either below $\Lambda_K$, where katoptrons are massive and decouple, or when fermions and katoptrons interact with distinct groups, as is the case for all the $U(1)$ and the $SU(5)$, $SU(5)^{\prime}$ couplings, couplings evolve according to
$c_N \equiv c(N, 6) = (11N-12)/6\pi$,  with $N=0$  for the $\alpha_{X}$, $\alpha^{\prime}_{X}$, $\alpha_{1}$, $\alpha^{\prime}_{1}$ couplings of  $U(1)_{X}$, $U(1)^{\prime}_{X}$, $U(1)_{1}$, $U(1)^{\prime}_{1}$ respectively, and $N= 2, 3, 5$ for the $SU(2)_{L}, SU(3)_{C}$ couplings and the $\alpha_{5}$, $\alpha_{5}^{\prime}$ couplings of $SU(5)$, $SU(5)^{\prime}$ respectively.
The relevant boundary conditions, noting that  $\alpha_K(\Lambda_K)$  $\sim   1$ and  $M_{Z} \sim 91.2$ GeV, are:
\vspace{-0,165cm}\begin{eqnarray*}
\alpha(\Lambda_{GUT})  &\equiv&  \alpha_{X}(\Lambda_{GUT})=\alpha_{5}(\Lambda_{GUT})=\alpha_{X}^{\prime}(\Lambda_{GUT})   =  
\alpha_{5}^{\prime}(\Lambda_{GUT})=\alpha_{K} (\Lambda_{GUT})\nonumber\\
&& (\rm{General}~\rm{ unification}~ \rm{ condition}) \nonumber \\
\alpha_2(\Lambda_{23})&=& \alpha_{3}(\Lambda_{23})
(\rm{Unification}~ \rm{condition}~ \rm{for}  ~
 SU(3)_{C} \times SU(2)_{L})
\nonumber \\
\alpha^{-1}_X(\Lambda_{23})&=&\frac{25}{24}\alpha_{1}^{-1}(\Lambda_{23})-\frac{1}{24}\alpha_{5}^{-1}(\Lambda_{23})
(\rm{Flipped}~  SU(5)  ~\rm{matching}~\rm{ condition})\nonumber\\
(\alpha_X^{\prime})^{-1}(\Lambda_{23})&=&\frac{25}{24}(\alpha_{1}^{\prime})^{-1}(\Lambda_{23})-\frac{1}{24}(\alpha_{5}^{\prime})^{-1}(\Lambda_{23}) 
 (\rm{Flipped}~  SU(5)^{\prime}  ~\rm{matching}~\rm{ condition})\nonumber\\
\alpha_1^{-1}&\equiv&\frac{3}{5}\alpha^{-1}_Y, (\alpha_1^{\prime})^{-1}\equiv\frac{3}{5}(\alpha^{\prime}_Y)^{-1}  
(\rm{Hypercharge}~ \rm{normalization})      \nonumber \\
 \alpha^{-1}_1(Mz) &=& 59.5, \alpha^{-1}_2(Mz) = 29.8, \alpha^{-1}_3(Mz)  = 8.5 
 ~( \rm{Experimental}~\rm{ input})
\label{boundary}
\end{eqnarray*}

These relations  yield $\Lambda_{GUT}$, $\Lambda_{23}$, and $\Lambda_{K}$, assuming a “big desert” between these scales and decoupling of the heavier degrees of freedom \cite{AppelCara}. Defining 
\begin{eqnarray*}
A &=&  1 - \frac{c_{3}-c_{2}}{\tilde{c_{3}}-\tilde{c_{2}}}, \hfill
 B =  \frac{\alpha_{2}^{-1}(M_{Z}) - \alpha_{3}^{-1}(M_{Z})}{\tilde{c_{3}}-\tilde{c_{2}}},  \nonumber \\
C &= & \frac{c_{2}+(1-A)(c_{5}-\tilde{c_{2}})}{c_{K}-c_{5}}, 
D =  \frac{\alpha_{2}^{-1}(M_{Z})  - B(c_{5}-{\tilde c_{2}})-1}{c_{K}-c_{5}},\nonumber\\
E&= &-\frac{ D(c_{K}-c_{0})-\frac{1}{24}\Bigl(B(c_{0}-\tilde{c_{2}})+25\alpha_{1}^{-1}(M_{Z})-\alpha_{2}^{-1}(M_{Z})\Bigr)+1}{C(c_{K}-c_{0})-\frac{1}{24}\Bigl(A(c_{0}-{\tilde c}_{2})+{\tilde c}_{2}-c_{2}\Bigr)-c_{0}}
\end{eqnarray*}

\noindent and using the boundary conditions listed above, we find 
\begin{eqnarray*}
\Lambda _{GUT}&=& M_{Z}\exp{\Bigl(E(1+C)+D\Bigr)} = 
M_{Z}\times 10^{17} \sim 10^{19}~  \rm{GeV} \sim M_{\rm{Planck}}\nonumber\\
\Lambda_{23}&=& M_{Z}\exp{\Bigl(AE+B\Bigr)} = 
M_{Z}\times 6 \times 10^{15} \sim 5 \times 10^{17}~ \rm{GeV}, \rm{and} \nonumber\\
\Lambda_{K}&=& M_{Z}\exp{E} = 
M_{Z}\times 11 \sim 1 ~ \rm{TeV}
\end{eqnarray*}
\noindent where $\Lambda_{23}=M_{Z}\exp{B}$ due to $A=0$. 
In addition, we find $\alpha(\Lambda_{GUT}) \sim 0.029$ and $\alpha_{3}(\Lambda_{23}) \sim 0.036$. A dynamical justification of the value of the unification coupling is attempted in the next section. Moreover, one observes that
\begin{equation}
\Lambda_{K} \sim  M_{\rm{Planck }} \exp{\Bigl(-\frac{6\pi}{17\alpha(\Lambda_{GUT})}\Bigr)},
\label{hierarchy}
\end{equation}

\noindent a relation rendering transparent the dynamical solution of the hierarchy problem due to katoptrons. The value of a single parameter remains then to be justified more fundamentally, {\it i.e.} $\alpha(\Lambda_{GUT})$, if a relation of the form $\Lambda_{23} \sim M_{\rm{Planck} } ~\alpha_{3}(\Lambda_{23})$  can be produced from the dynamics. 
Furthermore, the value of $\Lambda_{23}$ renders this scenario   safe, at first glance, with regards to proton decay.
\begin{figure}
\centering
\includegraphics[width=1\textwidth]{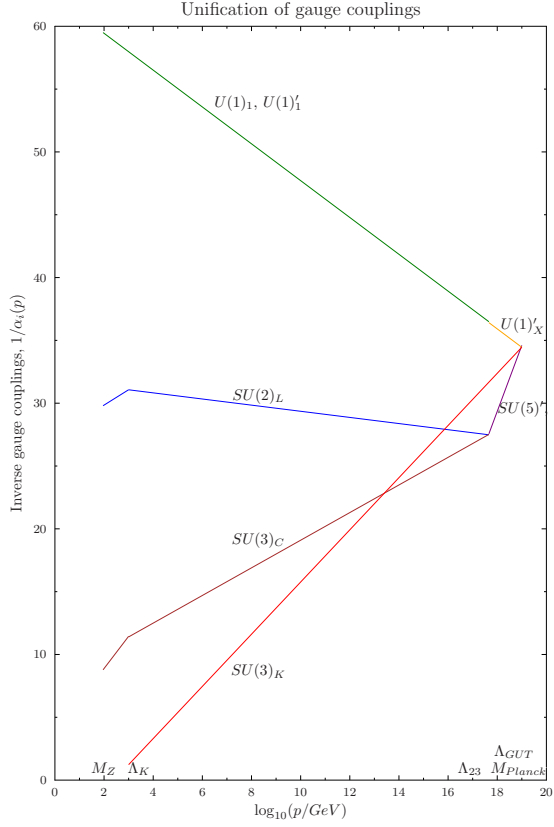}
\vspace{-2cm}
\caption{The renormalization of the gauge couplings at one loop}
\label{fig:test} 
\end{figure}

It has to be emphasized that if katoptrons kept interacting with the same hypercharge interaction 
as ordinary fermions, coupling unification would be impossible due to the faster running of the 
$U(1)_{1}$ coupling, as is obvious by the slopes of Figure 1. This problem was obviated in 
\cite{Triantap1} by the appearance of a relatively low scale of Pati-Salam symmetry breaking, something not 
possible here. In addition, inspection of Figure 1 shows that had katoptrons kept their own distinct 
$SU(3)^{\prime}_{C}$ and $SU(2)^{\prime}_{L}$ interactions,  coupling unification would 
require an EW symmetry breaking scale $\Lambda_{K}$ that would be too low. Furthermore, in
 such a case the scale $\Lambda_{23}$ would be too low to avoid an observable proton-decay 
rate. On the other hand, adding artificially additional matter in the present theory to slow down 
the running of the $SU(3)_{C}$ and $SU(2)_{L}$ couplings to enable them to unify  with an 
hypothetical  common $U(1)_{Y}$ coupling would require $\Lambda_{K}$ to be too high for EW 
symmetry breaking. 
The value of $\Lambda_{GUT}$ is close to $M_{\rm{Planck}}$ (see Figure 1), a result due to the slower running of some of the gauge couplings, since above $\Lambda_K$ both fermions and katoptrons contribute to the renormalization of the $SU(2)_{L}$ and $SU(3)_{C}$ couplings. This is also due to the use of the $SU(5)\times U(1)_X$ group and its primed partner. 
It is not expected that these results change significantly by higher-order calculations or by including contributions of the  fields responsible for symmetry breaking.

\subsection{Embedding within a larger symmetry}
Next, the symmetry breaking chain from a group $G$ down to $SU(5) \times U(1)_X \times SU(5)^{\prime} \times U(1)^{\prime}_X \times SU(3)_K$ and then down to the SM is explored using \cite{Slansky}. Non-zero vacuum expectation values (vevs) of effective composite fields are taken to lead to the breaking channels needed. These are fermion condensates arising non-perturbatively to safeguard gauge invariance at tree level.
 Coupling unification forces us to keep a distinct $U(1)_{1}^{\prime}$ for katoptrons breaking at $\Lambda_{K}$ in a way consistent with EW radiative corrections.  Therefore, apart from  fermion condensates of the form 
$<\bar{K}K>$, gauge-invariant 4-fermion condensates $<\bar{K}K\bar{K}F>$ are assumed to break the two hypercharge symmetries down to their diagonal subgroup $U(1)_Y$ and then together with $SU(2)_{L}$ down to $U(1)_{em}$ near the weak scale. Such operators arise again non-perturbatively  and are also needed to ``feed" mass to fermions \cite{Triantap2}.

 In order to have one generation of fermions in the $\bar{\textbf{5}}$ and $\textbf{10}$ of $SU(5)$, in addition to the right-handed neutrino, needed for the SM particles  to fit inside $SU(6)$ in an anomaly-free way (and similarly for katoptrons in $SU(6)^{\prime}$), we assign fermions in  $\bar{\textbf{6}} \oplus \bar{\textbf{6}}\oplus \textbf{15}$ \cite{SU6unif}. One generation of fermions fits inside 2 copies of 
$\bar{\textbf{6}} = (\bar{\textbf{5}},1) \oplus (\textbf{1},-5)$ and 1 copy of 
$\textbf{15} = (\textbf{5},-2) \oplus (\textbf{10},1)$. A $\bar{\textbf{5}}$ pairs up with a $\textbf{5}$ under $SU(5)$ acquiring thus GUT-scale masses. We are thus left with $(\bar{\textbf{5}},1) \oplus (\textbf{1},-5)  \oplus (\textbf{10},1)$ under $SU(5)\times U(1)_{X}$, i.e. a full fermion generation, plus a neutral lepton for each generation, and similarly for katoptrons. Assuming that effective fields in the \textbf{35} of $SU(6)$ and $SU(6)^{\prime}$  acquire non-zero vevs at $\Lambda_{GUT}$, $SU(6)\times SU(6)^{\prime}$ breaks down to $SU(5) \times U(1)_X \times SU(5)^{\prime} \times U(1)^{\prime}_X$.  
Note that this is not the only possible breaking channel for each $SU(6)$ left invariant by 
\textbf{35}, $SU(4) \times SU(2) \times U(1)$ and $SU(3) \times SU(3) \times U(1)$ being other 
examples. These are however phenomenologically unacceptable, since in these cases quarks would
not carry $SU(2)_L$ charges. Therefore,  it has to be proven that the "flipped" $SU(5)$ is the 
preferred symmetry-breaking channel. 

With regards to the breaking of $SU(5) \times U(1)_{X}\times SU(5)^{\prime} \times U(1)^{\prime}_{X}$ down to  $SU(3)_{C} \times SU(2)_L \times U(1)_1  \times U(1)_{1}^{\prime}$ at $\Lambda_{23}$, we use the fact that the $\textbf{35}$ of $SU(6)$   contains a $(\textbf{24},0)$ under $SU(5)\times U(1)_{X}$ (and similarly for the primed groups). 
Assuming that fermion composite operators with quantum numbers $(\textbf{1,24})$ and $(\textbf{24,1})$ under $SU(5)\times SU(5)^{\prime}$  acquire non-zero vevs near $\Lambda_{23}$, they break the two $SU(5)$ symmetries down to their $SU(3)_{C}\times SU(2)_{L}\times U(1)_{\tilde{1}}$ subgroups, while $U(1)_{X}$, $U(1)_{X}^{\prime}$ are left intact. In order to break the two resulting $SU(3)_{C}\times SU(2)_{L}\times U(1)_{\tilde{1}}$ symmetries down to their diagonal subgroup, since fermions and katoptrons interact with the same $SU(3)_{C}$ and $SU(2)_{L}$ interactions, it is necessary to couple somehow the -initially decoupled- left-handed and right-handed sectors of the theory in a way that the two $SU(5)$ groups break down to their diagonal subgroup. In the following, we assume that this coupling takes place, noting that a possible solution to this problem can be better justified within the context of the next subsection where it is presented.

Further breaking to the SM takes place by additional effective fields transforming as a $\textbf{20}$ of $SU(6)$, and similarly for $SU(6)^{\prime}$, which decomposes under $SU(5) \times U(1)_X$ like 
$\textbf{20} = (\textbf{10},1) \oplus (\textbf{10},-1)$. Near the $SU(5)\times SU(5)^{\prime}$  breaking scale $\Lambda_{23}$, these fields are taken to acquire  non-zero vevs and also break $SU(5) \times U(1)_X \times SU(5)^{\prime} \times U(1)_X^{\prime}$ down to  $SU(3)_C \times SU(2)_L \times U(1)_1 \times U(1)_{1}^{\prime}$, mixing their $U(1)_{X}$, $U(1)^{\prime}_{X}$ charges with the charges of  the $U(1)_{\tilde{1}}$ and  $U(1)^{\prime}_{\tilde{1}}$ groups in the final hypercharge groups $U(1)_{1}$, $U(1)_{1}^{\prime}$. 
We will see shortly how the effective fields $\textbf{20}$ and $\textbf{35}$ in each of $SU(6)$ 
and $SU(6)^{\prime}$ might emerge naturally in this theory. Moreover, one has to demonstrate that 
the alternative breakings of $SU(5)$ to $SU(4)\times U(1)$, and similarly for their primed 
partners, is disfavoured. The reason why alternative embeddings of the 
$SU(5)\times U(1)_{X}$ groups into $SO(10)$ or a  Pati-Salam symmetry for instance are not 
considered will become clear in the following. 

Our next goal is to expand the larger symmetry $SU(6) \times SU(6)^{\prime} 
\times SU(3)_K$ at the energy scale $\Lambda_{GUT}$ further.  Apart from being responsible for 
the dynamical EW symmetry breaking, $SU(3)_K$ plays the role of a gauged generation group 
for katoptrons. In addition, it prohibits gauge invariant vector-like masses corresponding to
 fermion bilinear operators formed by combining the known fermions with katoptrons. 
Even though it would be nice to have a generation gauge symmetry for the known fermions as 
well, this is excluded by phenomenological reasons, i.e. by the absence of flavour-changing 
neutral currents. It is however reasonable to assume, for naturalness and symmetry reasons, 
that such a generation symmetry $SU(3)_F$ existed at higher energies only to be broken at lower 
energies. An initial symmetry group of the form $SU(6) \times SU(6)^{\prime} \times SU(3)_F 
\times SU(3)_K$ would nonetheless have rank 14. In principle, such a symmetry may be 
embedded within a larger symmetry to allow for unification. 

For that purpose, we consider $G=E^{}_8 \times E_{8}^{\prime}$. Known fermions fit in $SU(6) \times SU(3)_F$ embedded in $E_7 \subset E_8$, while katoptrons fit in $SU(6)^{\prime} \times SU(3)_K$ embedded in $E_{7}^{\prime}\subset E_{8}^{\prime}$. Fermions sit in the $\textbf{248}$ reps of the two $E_{8}$s, and the ones in the $E_{8}^{\prime}$ exhibit the opposite chirality from the ones in $E_8$. We further assume that $SU(3)_F$  breaks at $\Lambda_{GUT}$. 
 The $\textbf{56}$ and $\textbf{133}$ of $E_7$ transform under $SU(6) \times SU(3)$  like $\textbf{56} = (\bar{\textbf{6}},\bar{\textbf{3}}) \oplus (\textbf{6,3}) \oplus (\textbf{20,1})$ and  $\textbf{133} = (\textbf{15},\bar{\textbf{3}}) \oplus (\bar{\textbf{15}},\textbf{3}) \oplus (\textbf{35,1}) \oplus (\textbf{1,8})$. If fermions initially sit in these reps, vector-like particles contained in $(\textbf{20,1}), (\textbf{35,1})$ and $(\textbf{1,8})$ obtain GUT-scale masses. We then need 2 copies of fermions in the $\textbf{56}$ of $E_7$  to give us the 2 $\bar{\textbf{6}}$s, and 1 copy in the $\textbf{133}$ of $E_7$  containing the $\textbf{15}$ of $SU(6)$ described above. This content, replicated thrice, corresponds to the fermion generations. Similar considerations apply for the primed groups corresponding to katoptrons. We discuss later fermions transforming under the conjugate $(\textbf{6,3})$ and $(\bar{\textbf{15}},\textbf{3})$.

How do $E_7$ and $E_{7}^{\prime}$  break down to  $SU(6) \times SU(3)_F$ and $SU(6)^{\prime} \times SU(3)_K$? The \textbf{133} of $E_7$ contains the  $(\textbf{35,1})$ of $SU(6) \times SU(3)$ which is needed for the breaking of $SU(6) \times SU(6)^{\prime}$ down to $SU(3)_{C}\times SU(2)_{L} \times U(1)_{\tilde{1}}\times U(1)^{\prime}_{\tilde{1}} \times  U(1)_{X} \times U(1)^{\prime}_{X}$.  Similarly,  the $\textbf{56}$ of $E_7$, when decomposed under $SU(6) \times SU(3)$, contains the $(\textbf{20,1})$  needed for the breaking of $U(1)_{\tilde{1}}\times U(1)^{\prime}_{\tilde{1}} \times U(1)_X \times U(1)^{\prime}_{X}$ down to $U(1)_{1} \times U(1)^{\prime}$. Considering non-zero vevs of  effective fields in the $\textbf{56}$ and $\textbf{133}$  of $E_7$ and $E_{7}^{\prime}$  may lead to the breaking sequence needed, since not only do these break  $E_7$, $E_7^{\prime}$, but they also contain the $(\textbf{20,1})$ and $(\textbf{35,1})$  fields for the breaking of the symmetries to the SM. 
The embedding of $E^{}_7\times E_{7}^{\prime}$ in $E^{}_{8}\times E_{8}^{\prime}$ is now straightforward, with left-handed and right-handed fermions initially contained in the fundamental reps of $E_{8}$ and $E_{8}^{\prime}$ respectively. Noting that $E_7 \times SU(2) \subset E_8$, and that the $\textbf{248}$ of $E_8$ decomposes under $E_7 \times SU(2)$ like $\textbf{248} = (\textbf{133,1}) \oplus (\textbf{1,3}) \oplus (\textbf{56,2})$, we find 2 copies of fermions in the $\textbf{56}$ and 1 copy of the $\textbf{133}$ in each of the $E_8$ and $E_8^{\prime}$, assuming that the extra $SU(2)$, $SU(2)^{\prime}$ symmetries in both $E_8$ and $E_{8}^{\prime}$ are also broken. 

Regarding the breaking of $E_8$ and $E_8^{\prime}$ down to $E_7\times SU(2)$ and its primed copy, the symmetric tensor product of 2 fundamental reps of $E_{8}$ yields $\textbf{248}\times \textbf{248} = \textbf{1} \oplus \textbf{3875} \oplus \textbf{27000}$. 
 In fact, $E_7 \times SU(2)$ is the only symmetric subgroup of $E_8$, denoted sometimes in this context as $E_{8(-24)}$, left invariant by \textbf{3875}. If fields, like fermion bilinears, having such quantum numbers acquired non-zero vevs, they could provide us with the desired
symmetry-breaking channel for $E_8$ and $E_{8}^{\prime}$, assuming that breaking to a symmetric subgroup is preferred over other channels. Moreover, the \textbf{3875} of $E_8$ contains the $\textbf{56}$ and $\textbf{133}$ effective fields under $E_{7}$ that could lead to the symmetry breaking channels exposed above.
The $\textbf{3875}$ leaves $E_7 \times SU(2)$ invariant, but it is quite large, and it being formed  is hard to justify by MAC arguments. 
 The $\textbf{248}$ contains the \textbf{56} and \textbf{133} of $E_{7}$ as well, and the corresponding channel is also attractive \cite{Adler}, so having a non-zero vev of an effective field in this rep might also suit our purposes and be ultimately  responsible for the breaking chain described. Since adding fundamental scalars is avoided, a candidate for this field  is formed by an antisymmetric tensor product of two fermions in $\textbf{248}$, {\i.e.} $\textbf{248}\times \textbf{248} = \textbf{248}_a$, where the subscript $a$ refers to its antisymmetric nature. This field however is not a Lorentz scalar, a fact used later in this work. Moreover, it is easy to check that the breaking channel above is more attractive than alternative ones leading from $E_{8}$ down to $E_{6}\times SU(3)$ or $SU(5) \times SU(5)$.

Next, we inquire how  the generation symmetry of the ordinary fermions $SU(3)_F$ breaks. We relax one of our assumptions and take the coupling $g_F$ of $E_7$ to be much larger at $\Lambda_{GUT}$ than the coupling $g_K$ of $E_{7}^{\prime}$, something lying at the heart of parity asymmetry. The large value of $g_{F}$  is assumed to trigger the self-breaking of $SU(3)_F$ down to $SU(2)_F$, via a non-zero fermion bilinear vev transforming under the $\bar{\textbf{3}}$ of $SU(3)_F$ originating from $\textbf{3}\times \textbf{3} \rightarrow \bar{\textbf{3}} \oplus \textbf{6}$. This is further broken down to  $U(1)_F$ with a vev transforming under the adjoint of $SU(2)_F$ coming from  $\textbf{2}\times \textbf{2} \rightarrow \textbf{1}\oplus \textbf{3}$, assuming that the singlet of this channel does not determine the correct vacuum. The remaining $U(1)_F$  breaks when the fundamental reps of $SU(6)$ pair up with each other, with the $\textbf{6}$ of \textbf{133} pairing up with the $\bar{\textbf{6}}$ of \textbf{56} of $E_7$. 
This scenario is in principle consistent with the self-breaking of the mirror-fermion generation symmetry $SU(3)_K$ at $\Lambda_K \sim 1$ TeV, since both symmetries self-break when their gauge couplings become strong \cite{Triantap2}. To assure that this does not create problems with coupling unification, we use the relation
$g(\Lambda_{23}) = \frac{g_Fg_K}{\sqrt{g^{2}_F+g^{2 }_K}} \sim g_K (\Lambda_{23})$ 
holding for the ($\alpha_{5}$, $\alpha^{\prime}_5$) and ($\alpha_{X}$, $\alpha^{\prime}_X$) couplings due to the  breaking of their respective gauge groups at $\Lambda_{23}$ down to their diagonal subgroups, since $g_F>>g_K$. All fermions at $\Lambda_{23}$ are then left with common abelian and non-abelian couplings, {\it i.e.} the weakest ones.

To sum up, denoting as $LG$ the symmetry $SU(2) \times SU(2)^{\prime}$, we assume that the following symmetry breaking chain is obtained, triggered by the $\textbf{248}_a$ condensate and starting from 
$M_{\rm{Planck}} \sim  \Lambda_{GUT}$:
\begin{eqnarray}
E^{}_{8} \times E_{8}^{\prime}~ ( \rm{at }~ \Lambda_{GUT})  \rightarrow && \nonumber\\
E_{7} \times SU(2)  \times   E_{7}^{\prime}
 \times SU(2)^{\prime} ~( \rm{at }~ \Lambda_{GUT}) \rightarrow && \nonumber \\
SU(6) \times SU(3)_{F}  \times SU(6)^{\prime} \times SU(3)_{K} 
\times [SU(2) \times SU(2)^{\prime}]~ (\rm{at }~ \Lambda_{GUT}) \rightarrow && \nonumber \\
SU(5) \times U(1)_{X}\times SU(5)^{\prime} \times U(1)^{\prime}_{X} \times SU(3)_K 
 \times LG  ~(\rm{at }~ \Lambda_{23}) \rightarrow && \nonumber \\
SU(3)_{C} \times SU(2)_L \times U(1)_1  \times U(1)_{1}^{\prime}   \times SU(3)_K 
\times LG ~ (\rm{at}~  \Lambda_K)   \rightarrow &&  \nonumber \\
SU(3)_C \times U(1)_{em}  \times  LG 
\end{eqnarray}

\noindent We write down the broken  $LG$ and its $SU(2)$ components, as well as $E_{7}$, $E_{7}^{\prime}$,  $SU(3)_{F}$, $SU(6)$ and $SU(6)^{\prime}$ in order to render the breaking sequence more transparent. Further dynamical justification of the breaking chain above follows in the next section.

Next, we address the appearance of conjugate generations in the considerations above.  These  are sometimes referred to as “mirror families”, and  would in principle appear within each of the $E^{}_8$, $E_{8}^{\prime}$. However, we assume that “mirror fermions” are just coming from a second $E_8$, i.e. $E_8^{\prime}$. What happens to the conjugate generations within each of the $E^{}_{8}$, $E_{8}^{\prime}$? This issue first appeared when we stepped from $SU(6)$ to $E_7$.
 If these conjugate generations were the “mirror partners” of the rest of the fermions, i.e. 
related to each other via a parity transformation interchanging their chirality, they would 
probably pair up with them, acquiring unification-scale gauge-invariant vector-like masses and 
disappearing from the low energy spectrum. If there were a symmetry reason why this is
forbidden, they would anyway never have masses much above 1 TeV, since in that case they 
would raise the weak scale at unacceptable levels. But even in that case, they would either share 
a gauged generation symmetry with the known fermions, something which is experimentally 
excluded, or not have such a gauge symmetry, in which case our dynamical symmetry breaking 
scenario involving strongly-interacting mirror fermions would break down. 
We must therefore dispense of the conjugate copies within each of the $E^{}_8$, $E_8^{\prime}$ in a consistent manner. 
This problem was circumvented in \cite{Triantap4} by introducing extra dimensions, since in 10
 dimensions one may define Majorana-Weyl spinors and impose a chirality condition, identifying 
thus the two kinds of fermion generations. In the present 4-dimensional approach however
 this trick is inapplicable. 

The answer lies within the channel the groups $E_8$ and $E_{8}^{\prime}$ break \cite{Slansky}. The group $E_{8}$ has just 2 symmetric subgroups, $SO(16)$ and $E_7 \times SU(2)$. Charge conjugation and parity transformations on fermion reps in each of these subgroups are distinct,  flipping the sign of a different number of $E_8$ roots. When embedding $SO(16)$ in $E_8$, charge conjugation $C^{\prime}$ amounts to complex conjugation, i.e. for a fermion rep $\textbf{R}$  in $SU(6) \subset SO(16)$ for instance,
$C^{\prime}: \textbf{R} \rightarrow \bar{\textbf{R}}$ and 
$\textbf{6}_L \rightarrow \bar{\textbf{6}}_L$, the bar denoting complex conjugation, while parity transformation $P^{\prime}$ changes chirality with no complex conjugation, {\it i.e.} $P^{\prime}: \textbf{R} \rightarrow\textbf{R}$ and
$\textbf{6}_L \rightarrow \textbf{6}_R$.
Consequently, $C^{\prime}P^{\prime}$ transformations taking particles to their antiparticles are distinct from parity transformations. On the contrary, in the embedding of $SU(6) \subset E_7$ in $E_8$, charge conjugation $C$ leaves the fermion rep invariant, 
{\it i.e.} $C: \textbf{R} \rightarrow \textbf{R}$ and
$\textbf{6}_L \rightarrow\textbf{6}_L$,  while a parity transformation $P$ implies not only chirality change but charge conjugation as well, {\it i.e.} $P:  \textbf{R} \rightarrow \bar{\textbf{R}}$ and
$\textbf{6}_L \rightarrow \bar{\textbf{6}}_R$, so that
$CP: \textbf{6}_L \rightarrow \bar{\textbf{6}}_R$.
Parity and $CP$ transformations have an identical effect in the $E_7 \times SU(2)$ case. Consequently, conjugate families appearing in $E_7$ and $E_8$ here are taken to be the antiparticles of fermions, not their “mirror partners”.
The fact that these are  not observed in nature characterizes the ``baryon asymmetry of the Universe" and is a distinct problem from the one of ``mirror" fermions. 
 The reason for choosing this particular symmetry-breaking channel is now clear, since a breaking involving  $SO(10) \subset SO(16)$ leading to a Pati-Salam (PS) scenario  creates problems with conjugate generations. 
An alternative PS scenario could in principle still be recovered via a channel involving $E_7$ and $E_6$  \cite{Triantap4}, but it leaves no room for the symmetry $LG$, which is central to the discussion below. 

\subsection{Connection with spinor gravity}
To proceed, we argue that $SU(2) \times SU(2)^{\prime} \approx LG$ contains the Lorentz symmetry $SO(3,1)$ (up to discrete subgroups omitted here for the sake of simplicity), an assumption motivated by the proximity of $\Lambda_{GUT}$ to $M_{\rm{Planck}}$. With regards to the non-compact nature of the Lorentz symmetry, note that we are dealing with the complexified versions $(C)$ of the corresponding groups. Therefore, to be more explicit, one should write the $E_{8}$ decomposition above as $E_{7}(C) \times SL(2, C) \subset E_{8}(C)$  \cite{Garibaldi}. One then gets from the two $E_{8}$s the group $SO(4,C) \approx SU(2,C) \times SU(2,C)^{\prime}$ (since $SU(2,C) \approx SL(2,C)$) which has both $SO(4)$ and $SO(3,1)$ as subgroups. A relevant mechanism should obviously be provided in order to lead to the relevant symmetry breaking  and to  the observed signature in nature, which is discussed later. The $LG$ group is therefore assumed to contain the Lorentz group
and it is taken to be a global, not a local symmetry. We list some arguments supporting this assumption. First, spin-1 gauge bosons associated with a gauged $LG$  are neither observed nor expected. Second, there are no prohibitive phenomenological constraints enforcing a local Lorentz symmetry \cite{Wetterich}. Third, taking Lorentz symmetry to be a global symmetry spontaneously broken allows the identification of some of the corresponding Goldstone bosons with the usual gravitons \cite{Percacci}\cite{Kostelecky}. Fourth, considering $LG$ as an unbroken symmetry would not allow the pairing of the fermions sitting in the $\textbf{6}$ and $\bar{\textbf{6}}$ of $SU(6)$ needed  to obtain the required fermion content. Last, breaking the Lorentz symmetry via an antisymmetric fermion tensor product transforming like $\textbf{248}_a$ under $E_8$ is   needed in order to obtain the symmetry-breaking channel to the SM, as described below. In order to render the relation between  $LG$ and the Lorentz group consistent, one needs to equate the $SO(3,1)$ coupling with the gravitational coupling. For coupling unification to work, one should extend the previous parity-breaking relation $g_F>>g_K$ from the two $E_{7}$ to the two $E_{8}$ groups. After the breaking of $SU(2,C)\times SU(2,C)^{\prime}$ down to the Lorentz group, $SO(3,1)$ should be left with the $g_{K}$ coupling of $E_{8}^{\prime}$, {\it i.e.} the weakest one. 

Before continuing, we need to address the Coleman-Mandula theorem prohibiting the total symmetry from being a direct product of a local Lorentz group with a gauged symmetry \cite{Coleman}. 
We choose to follow \cite{Percacci}, claiming that our starting point is a topological symmetric phase, in which the metric is initially absent and no S matrix is defined. As described below, the metric appears only as the product of non-zero vevs in the  “Higgs” phase. When the metric appears, Lorentz symmetry is a global symmetry presenting no further problems.

What follows below is a rough, initial investigation on whether the picture just described could in principle be incorporated in models unifying the Lorentz with the gauge groups, without going deep into the intricacies of such approaches. Within a framework similar to the one in \cite{Percacci}, we consider a metric of the form 
$g_{\mu\nu} = E^{m}_{\mu}(x)E^{n}_{\nu}(x) \eta^{}_{mn}=  E^{m}_{\mu}(x)E^{}_{\nu m}(x)$, where $\mu, \nu = 0, ..., d$ are spacetime indices, $m, n = 0,..., d$ are indices corresponding to the “internal” Lorentz symmetry of a $d$-dimensional spacetime with $\eta_{mn}= \rm{diag}(-1,1,...,1)$, and 
\begin{eqnarray}
E^{m}_{\mu}(x) &=& <\tilde{E}^{m}_{\mu}(x)> \sim \delta^m_{\mu}  M^{}_{\rm{Planck}} ~~\rm{for} ~ \mu, m = 0,...,3 \nonumber \\
E^{m}_{\mu}(x) &=& <\tilde{E}^{m}_{\mu}(x)> \sim  0  ~~~~~~~~~~~~~~\rm{for}~  \mu, m = 4,...,d \
\label{orderparam1}
\end{eqnarray}

\noindent are soldering forms (vielbeins), i.e. vevs of operators $\tilde{E}^{m}_{\mu}(x)$ breaking the  Lorentz symmetry spontaneously. The order parameter of this transition is a 1-form, not a 0-form. Global Lorentz symmetry is preserved only under combined Lorentz transformations on the internal Lorentz ($m$) and the ordinary spacetime ($\mu$) indices. Symmetric fluctuations of such a metric around the Minkowski spacetime are expected to produce Goldstone bosons identified with gravitons \cite{Percacci} \cite{Wetterich}. In the spinor gravity approach  \cite{Wetterich}, these vevs have a dynamical origin since they are expressed as fermion bilinear operators:
\begin{equation}
\tilde{E}^{m}_{ \mu}(x) = \frac{i}{2} \{{ \bar{ \Psi}(x) \gamma^{m} \partial_{\mu} \Psi(x)-\partial_{\mu}\bar{\Psi}(x)\gamma^{m}\Psi(x)}\}
\label{orderparam}
\end{equation}

\noindent where $\gamma^{m}$ are Dirac matrices in d dimensions and $\Psi, {\bar \Psi}$  are Grassmann variables in the irreducible spinor rep of the $d$-dimensional Lorentz group.

A relevant partition function, effective action and effective potential can then be formally defined, a 
prerequisite being an anomaly-free functional measure ${\cal D}\Psi$ preserving Lorentz and 
diffeomorphism invariance. 
This is expected to lead in principle, in lowest order in the effective potential expansion, to equations similar to the ones of General Relativity \cite{Wetterich}, in a way that spacetime is not treated as background but is incorporated in the equations non-perturbatively. In such a picture, physical distances are induced by fermion correlation functions and the appearance of a metric is inherently quantum-mechanical.
Difficulties in quantizing gravity  would show up in possible gravitational anomalies and 
in regularizing the corresponding effective potential. Although the finite number of counterterms in this context is encouraging for renormalizability, we do not pursue further the highly non-trivial issues arising in this setting.

The approaches  quoted above use large orthogonal groups in which the embedding of the Lorentz group takes place, 
treating Lorentz and gauge transformations in a unified manner as rotations in a
 higher-dimensional space (14 total dimensions in \cite{Percacci} corresponding to $SO(3,11)$, 16 
total dimensions in \cite{Wetterich} corresponding to $SO(16)$). The interpretation of these 
extra dimensions differs slightly in the two approaches. In \cite{Wetterich}, gauge symmetry 
arises by compactification of extra space dimensions. Therefore, in a complete approach we have to
replace ordinary derivatives by covariant ones in the quantities appearing in the
 four-dimensional $S_{eff}$, while this is not necessary for the higher-dimensional $S_{f}$. 
 On the other hand, extra dimensions in
 \cite{Percacci} are considered as internal dimensions corresponding to a unifying orthogonal 
group.  The two approaches are similar, with the action proposed in \cite{Percacci} corresponding to the 1-loop  effective action of \cite{Wetterich}.  This correspondence is realized when spacetime derivatives in 4d appearing in \cite{Wetterich} are replaced by gauge covariant derivatives, and fermion bilinears are treated as effective fields. 

The problem with large orthogonal groups however, as we saw above, is the emergence of conjugate generations which cannot be considered as anti-generations, but are just mirror copies of ordinary fermions. 
 To solve this problem and make connection with our model, we extend the unification symmetry to $E^{}_8\times E_8^{\prime}$, adopting the fermion-bilinear approach for the soldering forms  \cite{Wetterich} in order to maintain the dynamical interpretation of the breaking of $G=E^{}_8\times E_8^{\prime}$, where internal dimensions are connected with the appearance of gauge symmetries in 4d. We take some first steps exploring whether our model could give us a mechanism in principle compatible with such a dynamical metric-generation scenario. In fact, non-zero vevs of antisymmetric fermion bilinears sitting in the $\textbf{248}_a$ of $E_8$  and $E_{8}^{\prime}$ might lead to the breaking sequence needed.  The $SU(2)$ triplets in $\textbf{(1,3)}$ and doublets in $\textbf{(56,2)}$ contained in the decompositions of the $\textbf{248}_a$  of each of the $E_8$, $E_8^{\prime}$ under $E_7\times SU(2)$ can break spontaneously  $LG$  after acquiring non-zero vevs, leading thus to a dynamically generated metric tensor by condensates, assuming that $<\bar{\Psi}\gamma^{m} \partial_{\mu}\Psi> = 0$ for $m>3$. 
  The vevs of fields in the $\textbf{(1,3)}$ of $E_7\times SU(2)$ and $E_7^{\prime}\times SU(2)^{\prime}$ constitute the antisymmetric MAC of $E^{}_8\times E_{8}^{\prime}$ breaking down to $E^{}_7 \times E_{7}^{\prime}$,  possibly justifying their dominance in determining spacetime dimensionality over the $\textbf{(56,2)}$ and $\textbf{(133,1)}$ vevs breaking subsequently $E^{}_7 \times E_7^{\prime}$ and the remnant of $LG$.

 In order for $LG$ to break down to its diagonal subgroup $SO(3,1)$, one could consider vevs of 4-fermion composite operators, corresponding to gauge-invariant terms in the initial action and transforming as $\textbf{(1,3,1,3)}$ under
 $E_7\times SU(2) \times E_7^{\prime}\times SU(2)^{\prime}$. In such a case, the interpretation of the metric would arise from an approximation of the form $g_{\mu\nu} = <\tilde{E}^{m}_{\mu} \tilde{E}^{}_{\nu m}> \sim <\tilde{E}^{m}_{\mu}>< \tilde{E}_{\nu m}>$. One way to couple the two initial -in principle decoupled- $E_{8}$ sectors in a way having the same end result as such operators is to consider an initial symmetry of the form 
$E^{}_{8}\times E^{\prime}_{8} \times \Bigl(Z^{E_{8}}_{2}/Z^{SU(2)}_{2}\Bigr)$, where the discrete $Z_{2}$ symmetries interchange the two $E_{8}$ groups and their $SU(2)$ subgroups respectively, in a construction similar to \cite{Triantap4}\cite{Forgacs}. A deeper analysis of this scenario might have the potential of not only explaining the dimensionality of our space-time, but its signature as well, i.e. why the group $SO(3,1)$ emerges instead of $SO(4)$ 
or $SO(2,2)$ for instance. In any case, it is quite interesting to note that, in this picture, the same type of fermion condensate is responsible not only for metric generation by for gauge symmetry breaking as well.

The next step is to consider the effective fermionic action in 4d in a derivative and fermion-field expansion  to 1-loop taking a unified form similar to the one in \cite{Wetterich}
\begin{eqnarray}
S_{eff}& \sim & \int d^{4}x \det{}\Bigl(E^{m}_{\mu}(x)\Bigr) \Bigl(c_{1}+c_{2}R+
 c_{3}\bar{\Psi}(x)\gamma^{m}E_{m}^{\mu}(x)D^{}_{\mu}\Psi(x) + ... \Bigr)
\label{effaction}
\end{eqnarray}

 \noindent plus gauge kinetic terms, where  $\det{}\Bigl(E^{m}_{\mu}(x)\Bigr)$ $=\sqrt{\det{}(-g_{\mu\nu})}$ $\neq 0$ while $\mu, \nu, m = 0, ...,3$, $c_{1,2,3}$ are constants and   $D_{\mu}$ is the gauge-covariant derivative corresponding to the $E_{7}\times E_{7}^{\prime}$ gauge symmetry after compactification, in a process assumed to respect Lorentz and 
diffeomorphism invariance. This symmetry should probably be corrected by a multiplicative factor of the form $\Bigl(Z^{E_{7}}_{2}/Z^{SU(5)}_{2}\Bigr)$ interchanging the two $E_{7}$ groups and their respective $SU(5)$ subgroups in a way that couples the right- and left-handed sectors of the theory during the compactification process and causes the breaking of $SU(5) \times SU(5)^{\prime}$ to the diagonal subgroup discussed in the previous subsection. The first term of the action above gives a cosmological constant $\Lambda$, the second the Ricci curvature, and the third the action in \cite{Percacci}. 
The result above is expected to stem in principle from an  action containing the expression
\begin{equation}
S_{f}\sim \int d^{d}x \det{}(\tilde{E}_{\mu}^{m})
\label{action}
\end{equation}

\noindent in $d$ dimensions, while other invariants give rise to higher order terms. Details on the compactification mechanism shedding light on questions like ``why are other dimensions left compactified after the $E_{7}\times E_{7}^{\prime}$ symmetry breaking?" are left for future studies.

 Consider now a 2-form $\tilde{G}_{mn}$ as a generalized, internal ``metric" defined on a $d$-dimensional manifold expressed as
$\tilde{G}_{mn} = \tilde{E}_{m}^{p}\tilde{E}^{}_{np}$, with $m, n, p$ spanning all $d$ dimensions and $\tilde{E}_{m}^{p}$ defined as before. Non-zero vevs sitting in the $\textbf{248}_{a}$ of $E_8$ and $E_8^{\prime}$ give rise to a breaking of the initial symmetry down to $E^{}_7\times E_{7}^{\prime}$ and to an effective metric
$\tilde{G}_{eff}$. In lowest order, after splitting the coordinates to spacetime  ($x$) and internal ($y$) ones, this should take the form
\begin{equation}
\tilde{G}_{eff}=\left( \begin{array}{cc}
g^{}_{\mu \nu} +g^{}_{kl}A_{\mu}^{k}A_{\nu}^{l} & g^{}_{kl}A_{\mu}^{k}\omega^{l}_{b}  \\
g^{}_{kl}A_{\nu}^{k}\omega^{l}_{a} & g^{}_{kl}\omega^{k}_a \omega^{l}_b \\ 
  \end{array}\right) 
\end{equation}
\noindent where the $\mu, \nu$ indices refer to $x$- while the $a, b, k, l$ indices to $y$-coordinates.  
Assuming that lower-dimensional fields are independent of $y$ and imposing local invariance of $\tilde{G}_{eff}$ under diffeomorphisms, we identify  the 4d  effective fields as $g^{}_{\mu \nu}(x) = E^{m}_{\mu}(x)E^{}_{\nu m}(x)$  (Riemannian metric), 
$A_{\mu}^{a}(x)$ (Spin-1 Kaluza-Klein $E^{}_{7}\times E_7^{\prime}$ fields) and 
$g^{}_{kl}(y)\omega^{k}_a \omega^{l}_b$ (Spin-0 fields, metric of internal dimensions).
The Maurer-Cartan 1-forms on the internal coordinates $y^b$ defined as $\omega^a \equiv \omega^{a}_{b}dy^{b}$ are dual to the Killing vectors $K^{}_b \equiv K_{b}^{a}\partial^{}_{a}$ satisfying the commutation relations $[K_a, K_b] = c^k_{ab}K^{}_{k}$ with $c^k_{ab}$  the $E_7\times E_7^{\prime}$ structure constants, i.e. $\omega^{a}K^{}_b = \delta^{a}_b$. Obviously, Killing vectors preserve the metric of the internal dimensions under the transformation of the coordinates $y^a \rightarrow y^a + \epsilon^{b}(x)K^a_{b}$, with $\epsilon^{b}(x)$ arbitrary functions of $x$. These transformations correspond to isometries of the internal manifold expressed by the $E_7\times E_7^{\prime}$ algebra.  It is important to realise that in this setting, gauge fields are conceptually associated not only with isometries of an internal manifold, but also with appropriate fermion two-point correlation functions.

We then investigate the dimensionality $d$ of space to integrate our Lagrangian over. Since spacetime and internal dimensions are treated on an equal footing, naturalness reasons lead us to consider the action as an integral over a manifold having the isometry $E^{}_8 \times E_8^{\prime}$ up to discrete factors which we omit in the following discussion for the sake of simplicity. The proper number of (complex) dimensions to integrate our Lagrangian over is then $d=16$, equal to the number of roots of the groups involved.  Integrating over the 14 extra internal  dimensions 
gives us $E_7 \times E_7^{\prime}$, 
which by the way correspond to 2 lattices formed by the unit-norm imaginary Caley octonions, and
should leave us with 2 complex, i.e. 4 real, ordinary spacetime dimensions.
The internal dimensions are then assumed to be compactified at a size of around $1/M_{\rm{Planck}}$ to avoid the appearance of Kaluza-Klein excited states at energies lower than $M_{\rm{Planck}}$.
We do not study effects arising from 0-spin fields connected to the compactification  process such as the dilaton. The properties of a 4d effective potential formed by such scalars associated with the ``shape" of the compactification space might shed light on $\Lambda$, its relative size to the root norm $1/M_{\rm{Planck}}$ and other cosmological issues like inflation. In any case, these are  expected to decouple at lower energies since they have $M_{\rm{Planck}}$ masses.

It is conjectured below that the  space with the isometries needed is the quotient space of the 16d maximal torus $T^{16}$ by  the lattice $\pi \Gamma_8 \times \pi \Gamma_8$ generated by the roots of $E_8 \times E_8^{\prime}$ multiplied by integer multiples of $\pi$. Allowed coordinate transformations take the form $\delta y^a \in \pi \Gamma_8 \times \pi \Gamma_8$, {\it i.e.}
$\delta y^{a} = \pi \sum_b m^{}_{b}e_{b}^{a}
\label{lattice}$, with $e_{b}$ a basis of the root space and $m_{b}$ integers. Roots can be associated with the Killing vectors $K_{b}$ giving rise to the metric isometries. The  conserved  conjugate momenta $p_b \sim i K_b$ corresponding to Killing vectors are associated with the same lattice and can be expressed as $p_{a} =  \sum_b m^{\prime}_{b} e^{b}_{a}
\label{mom1}$, where $m^{\prime}_{b}$ are again integers. 
Invariance requirements for the metric and single-valuedness of  the plane wave $e^{i2p_{b}y^{b}}$ after coordinate transformations  imply that $\sum_a p_a \delta y^{a} = \pi n, n \in {\bf Z}$, which requires that
$p_{a} =  \sum_b {\tilde m}^{}_{b} {\tilde e}^{}_{ba}$, with ${\tilde e}_{b}$ a basis of a space dual to $e^{b}$, {\it i.e.} $\sum_c e^c_{a}{\tilde e^{}_{cb}}=\delta_{ab}$, and ${\tilde m}_b$ integers. Simultaneous validity of the expressions for the momenta above can only be achieved if the underlying lattice is even and self-dual, like the lattice $\Gamma_8 \times \Gamma_8$. Such lattices exist only in $d=0 \hspace{1mm}(\rm{mod}~  8)$ dimensions. When  $d=16$, the only such lattices are $\Gamma_8 \times \Gamma_8$ and  $\Gamma_{16}$. Note that the same space has already been used in the context of heterotic string theory, the crucial difference being the origin of the quantization condition; instead of considering
strings on a continuous background, the idea here is to work on a discrete space from the start. 
 After symmetry breaking, $LG$ should carry traces of this discrete structure, while taking the root norm to be of size $1/M_{\rm{Planck}}$ is expected to approximate satisfactorily the continuum at low energies. More work is obviously needed in order to check rigorously the validity of this construction, which will be quite useful in the next section.

In the following, an effort is made to motivate further our choice of $E^{}_8\times E_8^{\prime}$ as a unification symmetry group, apart from the self-duality feature which might prove to be unique and crucial for unifying spacetime with gauge symmetries. The arguments presented below might reduce the arbitrary nature of such a choice. The appearance of this symmetry might be due to a relevant phase transition in a discrete space and might be related to the fact that each of the two 8d $E_8$ lattices, which are also even and self-dual, offer the densest sphere-packing and correspond to the highest ``kissing number" configuration known in 8d. This property allows by the way their use in other scientific areas like coding theory, since they offer the most efficient information transmission \cite{Conway}.  To use this result in the present context however, the 8d spaces of the two $E_8$ groups have to be treated distinctly, something consistent with the parity violating assumption $g_F >> g_K$ made previously, since couplings emerge from the volumes Vol$_{c}$ of the compactified dimensions and  are inversely proportional to them.
Taking the $E^{}_8$, $E_8^{\prime}$ to be localized on distinct 8d hypersurfaces (hyps) and the compactification radius corresponding to $E_{8}$ to be much smaller than the one of $E_{8}^{\prime}$, {\i.e.}
Vol$_c(E_8$ hyps$) <<$ Vol$_c(E^{\prime}_8$  hyps$)$, could possibly give a geometrical interpretation of parity-symmetry breaking. 
Treating the two $E_8$s distinctly from this particular viewpoint is anyway required, since, while each of the two $E_8$ lattices provide the highest kissing number and densest sphere packing in 8d, in 16d other lattices like the Barnes-Wall (BW) lattice provide a higher kissing number and denser sphere packing than $\Gamma_8 \times \Gamma_8$, even though they are not self-dual. In particular, the $\Gamma_8 \times \Gamma_8$ kissing number equals 480, while the BW lattice has a kissing number equal to 4,320. However, since lattices like BW do not correspond to any root system, they cannot generate the symmetries needed. 

It is very important to note that densest sphere packings  in higher dimensions $d>8$ are most likely either disordered, not corresponding to lattices, or the lattices they correspond to are not associated with root systems and thus known gauge symmetries. The disordered packing phenomenon appears already when $d=9$
and it is due to the fact that the packing density $\phi$, defined as the ratio of volume of one 
sphere to the volume of the  corresponding lattice's fundamental cell, is falling exponentially with $d$. 
The $E_8$ lattice $\Gamma_{8}$ for instance provides a packing density of only 
$\phi(\Gamma_{8}) = \pi^{4}/384 \sim 25\%$
 in eight dimensions, while in three dimensions, the highest sphere packing density, in an 
arrangement correctly conjectured by Kepler and being part of Hilbert's 18th problem, reaches 
around $74\%$.  
However, even when such higher-dimensional densest sphere packings correspond to lattices, like the ``Leech" lattice in 24d, these do not correspond directly to a Lie group's root system. Therefore, all other cases for $d>8$ cannot easily lead to the symmetries needed for phenomenology and unification considerations.

An additional argument supporting this scenario is related to the concept of optimal lattices.  In order for $\Gamma_8 \times \Gamma_8$ to arise naturally, it could be shown that it either extremizes an effective ``potential" between lattice points, or equivalently that nature is based on a new fundamental principle reformulating the ``least-action" principle,  requiring ``most efficient information transmission" for each of the 2 lattices of the $E_8$ groups, justifying thus their choice as the vacuum of our world.  This lattice would then have to be universally optimal, independently of the specific form of the ``potential" used, apart from general requirements of being repulsive at short- and attractive at long distances. Note that universal optimality of this lattice has already been investigated and is true  for  potentials,  as functions of the distance between the centres of the spheres located at the origin of the roots, which are decreasing monotonically fast enough \cite{Universal}.
Repulsion at short distances amounts to the impossibility of having 2 lattice sites occupying the same position, which is equivalent to Pauli's exclusion principle. Attraction at large distances leading to the formation of this lattice is achievable {\it via} dynamics favouring long-range order.  
 
Alternative justification for the number of internal dimensions (14 here) as multiple of seven might come from arguments based again on densest sphere packing and involving optimal vacuum energy density \cite{Pushkin}.   In any case, all these arguments might be an indication that we are approaching a theory starting from an action as simple as the one in equation \ref{action} and yielding the dimension of the internal space and its isometries by the dynamics of the action itself, without having to postulate them {\it a priori}. This is similar in spirit with emergent geometry and gravity \cite{Markopoulou}, although it seems more general since it tries to include all the gauge symmetries as well. The next section deals with this issue in more detail, trying to crystallize these thoughts in a more concrete way.

\section{Critical behaviour and emergence of symmetry}
\label{sec:5}

\subsection{Estimating the unification coupling}
In order to connect our action to the results above, one needs first to justify the value of  $\alpha(\Lambda_{GUT})$ calculated  in section 2. We are dealing with a critical phenomenon  breaking  $E^{}_{8} \times E_{8}^{\prime}$ and  having as order parameters the non-zero vevs  in equations \ref{orderparam1} and \ref{orderparam}. The relevant critical parameters are the couplings of the two $E_8$s.  
The order parameters are assumed here to scale as 
$E^{m}_{\mu} \sim \frac{p^{m}p_{\mu}}{M_{\rm{Planck}}}<{\bar \Psi}\Psi>$
with fermions $\Psi$  in the fundamental reps of $E_{8}$ and $E_{8}^{\prime}$ and $|p^m|, |p_{\mu}| \sim M_{\rm{Planck}}$. The $<{\bar \Psi}\Psi>$ condensate, corresponding to the MAC, is assumed here to be the catalyst for the formation of the antisymmetric condensates in the $\textbf{248}_a$  of the two $E_{8}$s. Similarly, it is taken to be the catalyst for the formation of the 4-fermion operators discussed in section 2. This should obviate problems arising from the fact that the antisymmetric channel is by itself not as attractive as the singlet one. This implies that the values of the critical couplings for the formation of such vevs are at least close to each other. We return to this issue in the next subsection.

Since $E^{}_8 \times E_8^{\prime}$ breaks at around $M_{\rm{Planck}}$, we first estimate this coupling by using the Nambu-Jona-Lasinio (NJL) formalism, which  in 4d gives
\begin{equation}
\frac{<{\bar \Psi}\Psi>}{M_{\rm{Planck}}^{2}} \equiv m = \frac{\lambda}{M_{\rm{Planck}}^{2}} \int_{0}^{M_{\rm{Planck}}^{2}}\tilde{k}^{2}d\tilde{k}^{2}\frac{m}{\tilde{k}^{2}+m^{2}}
\label{NJL}
\end{equation}

\noindent with fermion mass $m$ and  an effective coupling $\lambda$ given by 
\begin{equation}
\lambda = \Biggl(1-\frac{m^{2}}{M_{\rm{Planck}}^{2}}\ln{\left(1+\frac{M_{\rm{Planck}}^{2}}{m^{2}}\right)}\Biggr)^{-1}
\label{NJL2}
\end{equation}

\noindent and having a critical value $\lambda_{c}=1$ (see for instance \cite{critical}). Assuming that $\lambda$ is determined by the value $a_{SB}$ of the gauge coupling at the symmetry breaking scale, we find
$\lambda \sim \frac{3\alpha_{SB}}{4\pi}C_{2}$ with $C_{2}= 30$  the quadratic Casimir invariant of $E_{8}$ \cite{Shel}. 
A value  $m \sim M_{\rm{Planck}}/10$  yields $\lambda \sim 1.05 > \lambda_c$ and $\alpha_{SB} \sim 0.15$. Although this number is small due to the magnitude of $C_{2}$, it is larger than $\alpha(\Lambda_{GUT}) = 0.029$ found in section 2. In fact, $\alpha(\Lambda_{GUT})$ is close to its critical value \cite{critical} in the case of unbroken $E_{8}$ gauge symmetries. This may be due to the fact that gravitational interactions stemming from  $LG \subset E^{}_{8} \times E_{8}^{\prime}$ are associated with massless Goldstone bosons identified with gravitons, and are also connected to $E^{m}_{\mu}$ which  have infinite correlation length. This implies a graviton propagator whose scalar part for small momenta $k$ scales as $1/k^{2}$ \cite{Litim}. Since in a unified setting the gravitational constant scales as $G \sim \lambda/M^{2}_{\rm{Planck}}$, neglecting this infrared (IR) behaviour amounts to neglecting the masslessness of gravitons. The enhanced contribution to the condensate integral stemming from the IR region might be responsible for the smaller critical value of $\lambda$.

Although a full calculation requires a complete quantum theory of gravity, we attempt to estimate roughly this critical coupling by replacing  $1/M_{\rm{Planck}}^{2}$ above with a gauge-boson propagator $1/(k-\tilde{k})^{2}$, $k$ being the external momentum. After angular integration this expression is replaced by $1/max(k^{2},\tilde{k}^{2})$. Taking $m$  and $\lambda$ to be momentum-independent, we have
\begin{equation}
\frac{<{\bar \Psi}\Psi>}{M_{\rm{Planck}}^{2}} \equiv m \sim \lambda \int_{0}^{M_{\rm{Planck}}^{2}}\frac{\tilde{k}^{2}d\tilde{k}^{2}}{max(k^{2},\tilde{k}^{2})}\frac{m}{\tilde{k}^{2}+m^{2}}
\end{equation}

\noindent yielding $\lambda = \Bigl(1+\ln{(\rho^{2}+1)}-(1+m^{2}/k^{2})\ln{\left(1+k^{2}/m^{2}\right)}\Bigr)^{-1}$,
where $\rho \equiv M_{\rm{Planck}}/m$.  Naturalness arguments lead us to consider $1 \leq \rho \leq 10$. The regime $k \gg m$ is then equivalent to $k \sim M_{\rm{Planck}}$, in which case $\lambda$ is given by the same expression as in equation \ref{NJL2}, so $\lambda_{c}=1$. 
For $k \sim m$ however, $\lambda \sim 1/\bigl(1+\ln{(\frac{\rho^{2}+1}{4})}\bigr)$. Taking $\rho \sim 10$ as before, we have $\lambda \sim 0.24$. Then, we find $ \alpha_{SB} \sim ~C^{-1}_{2} \sim 0.03 \sim \alpha(\Lambda_{GUT})$,
 allowing thus a dynamical interpretation of $\alpha(\Lambda_{GUT})$ found in the previous section. Similar results are obtained for $k \ll m$. Criticality is not apparent for $k \leq m$ and $\rho \sim 10$, but $\alpha$ is close to $\alpha_{c} = \frac{\pi}{3}C^{-1}_{2} \sim 0.035$ predicted for self-energies $m=m(k)$ in unbroken gauge theories \cite{critical}. 
With regards to equation \ref{hierarchy}, the wide hierarchy between $M_{\rm{Planck}}$ and the weak scale can then be traced back to  the magnitude of $C_{2}$, since 
\begin{equation}
\Lambda_{K} \sim M_{\rm{Planck}}\exp(-1.23~C_{2}).
\end{equation}
The coupling in higher dimensions being  inversely proportional to Vol$_c$, a shrinking compactification space until the above equation is satisfied might provide a relevant geometrical interpretation. 

\subsection{A ``toy" model for symmetry emergence}
 As promised in the previous section, we now take a first glance at the dynamics which might lead to the emergence of symmetry in the first place by studying a relevant phase transition. Deferring rigorous justification of this approach to a more detailed future study, we apply techniques  borrowed from  similar studies in solid-state physics, chemistry, biology and even sociology. The common starting ground is the emergence of particle configurations exhibiting spontaneous self-organization transitions in ordered structures and nucleation, like DNA, neural networks or crystals, {\it i.e.} processes characterized by ``self-organised criticality" \cite{selforg}. 
For instance, a numerical study of a liquid-to-crystal freezing transition of hard spheres  shows that the emerging crystal, after the liquid has been subject ``slowly"  to pressure, minimizes its potential by a densest-sphere packing arrangement \cite{Mau}. Faster compression rates lead to crystals containing defects and to ``liquid-to-glass" transitions, corresponding here to a cosmological scenario lacking the symmetries needed, since glass corresponds to a disordered phase. 
According to these studies, effective-potential minimization on  the emerging lattices is carried
 over to the dual lattices, in a way that long-range, non-local effects in one lattice can be 
substituted by local effects in its dual. Therefore, the self-duality property of the $E^{}_8\times 
E_8^{\prime}$ lattice mentioned previously might provide a further advantage of the  scenario 
proposed here. 
First, we find the relevant universality class,  allowing us  to predict the qualitative behaviour of the system by studying a simpler model Hamiltonian belonging to the same class. One crucial factor determining universality class  is dimensionality, noting that 8, the dimension of the $E_{8}$ lattice, is the upper critical dimension for several random physical systems, generic lattice trees and some polymer and percolation models \cite{Slade}, appearing to be critical also for transitions in glasses \cite{Parisi} and allowing the application of mean-field theory results.

We place our action on a lattice to see if its qualitative behaviour can be inferred by simpler or similar systems in lattice gauge theories or solid-state systems. The effective action $S_{lat}$ stemming from equation \ref{effaction} to lowest order, apart from the Einstein-Hilbert terms, is written as $S_{lat} = \sum_{<i,j>}{\cal E}_{ij}\bar{\Psi}_{i}\Psi_{j}$,
where ${\cal E}_{ij}$ is an antisymmetric matrix proportional to the system's volume and  encoding information on $E^{m}_{\mu}$. The fact that the sum over the lattice sites $i, j$  is restricted over nearest neighbours, denoted by $<i,j>$, originates  from the partial derivative in the action. Gauge fields and spacetime do not appear yet,  our goal being
 to have them emerge rather than postulate their appearance {\it a priori}. In the process, we will question briefly whether
 the assumption that elementary particles are a manifestation of topological or vacancy defects within the emergent complex structure can be made plausible.

The form of $S_{lat}$ is reminiscent of the Edwards-Anderson (EA) spin-glass Hamiltonian with zero external field,  given by 
$H_{EA} = -\sum_{<i,j>}J_{ij}S_{i}S_{j}$ \cite{EA},
with locally-interacting spin fields $S_{i,j}$ taking values $\pm 1$ and $J_{ij}$ random variables obeying a zero-mean  distribution function. The minus sign in front of $H_{EA}$ makes the EA model favour long-range, ferromagnet-like, order at low temperatures $T$.  To render this  compatible with $S_{lat}$, we treat the fermion kinetic term as corresponding effectively to an interaction. 
A problem arises from the fact that the  $S_{i,j}$ are bosonic while the $\Psi_{i,j}$ are fermionic fields.  We assume in the following that the two models above belong to the same universality class in order  to draw some conclusions on their qualitative behaviour which should not be influenced by  field statistics, at least for large $T$. The anticommutativity of fermions,  an expression of Pauli's exclusion principle, guarantees that these lie on distinct locations, something which in bosonic Ising- or Potts-inspired models is ensured by assuming {\it a priori} an underlying lattice. Therefore, positing the existence of a lattice approximates one feature of Fermi statistics.

The universality class of similar models  depends on the average coordination number $c$, {\i.e.} the number of nearest neighbours of each lattice site. An 8d square lattice ${\cal Z}^{8}$ with $c=16$  gives a different behaviour from the $E_{8}$ lattice having $c=240$, even though they are both defined in 8d. Models with large $c$, such as the one defined on a $E_{8}$ lattice, exhibit dynamics described by the mean-field approximation \cite{Parisi2}\cite{Hellmund}. The EA model approaches thus models on random Bethe lattices and complex network theory studying random structures. These exhibit topological phase transitions when the bond concentration probability $p$ exceeds a critical value $p_{c}$ called ``percolation threshold", which for large $c$ is given by $p_{c} \sim 1/c$. For $p < p_{c}$, only finite clusters of edges connecting lattice sites appear, while for $p > p_{c}$ the whole lattice is occupied by a huge cluster \cite{Hellmund}.  This proves to be crucial for the discussion below.

In order to study the emergence of the $E_{8}$ lattices from first principles, we use a model suitable for percolation phenomena, a ``toy" model in our case, assumed here to belong to the same universality class, {\it i.e.} the single-state ($q=1$) Potts model, with Hamiltonian $H_{P}$, similar to $H_{EA}$, given by $H_{P} = -J\sum_{<i,j>}\delta(S_{i},S_{j})$, where $J>0$ is the coupling strength and $\delta(S_{i},S_{j})=1$ when $S_{i}=S_{j}=1$ and zero otherwise. The partition function $Z = \sum_{C_{i}}e^{-\beta H_{P}}$ is given by
\begin{equation}
Z = \sum_{C_{i}}Z_{i} = \sum_{C_{i}}\Bigl(e^{\beta J}-1\Bigr)^{E_{i}}
\label{partition}
\end{equation}

\noindent where the sum is over clusters $C_{i}$ consisting of $E_{i}$ edges and $1/\beta = k_{B}T$. Here, $p$ is given by $p = 1 - e^{-\beta J}$ and increases with decreasing $T$. This model exhibits a 2nd-order phase transition for low $T$, which for large $c$ implies a mean-field behaviour for the 2-point correlation function given by $<\tilde{S}_{i}\tilde{S}_{i+k}> \sim 1/k^{2}$. 
 Such discrete models are usually studied on lattices with  given dimensionality $d$ and  $c$.  In the following, we explore the behaviour of a system of nodes minimizing its free energy by adjusting its $d$ and $c$ in order to form an optimal lattice. 

For high $T$, 
$w \equiv e^{\beta J}-1\sim p \sim \beta J \ll 1$ in equation \ref{partition} implies that only clusters with few edges contribute significantly to the partition function. 
In a competition between annihilation and aggregation of large clusters, those of low $d$ and $c$ dominate. Filaments or low-dimensional surfaces formed by edges connected with each other are expected to form topological entities  resembling ``time" with 1 or 2 ``space" dimensions, without more structure.
 It is worth reminding however that such configurations are not accurately described by
 mean-field theory.
In this high-$T$ regime, each cluster $C_{i}$ consisting of $E_{i}$ edges gives a positive contribution to the system's free energy $F_{C_{i}} = - \ln{Z_{i}}/\beta \sim -E_{i}\ln{(\beta J)}/\beta$. The positive free energy, compensated by the system's gradual cooling, is identified with a vacuum energy, {\it i.e.} with $\Lambda$. It is accompanied, for a system of volume $Vol$, by a negative pressure $P=-F_{C_{i}}/(Vol)$ leading to expansion and probably to an inflationary scenario for the early Universe. The same cluster $C_{i}$ gives a negative contribution $S_{C_{i}}= -\partial F_{C_{i}}/\partial T \sim k_{B}E_{i}\Bigl(\ln{(\beta J)}-1\Bigr)$ to the system's entropy $S$, rendering the formation of large clusters of edges very costly, energetically and entropy-wise. This system of nodes, a rough model of the ``pre bib-bang" world, lies initially in a highly-probable state, {\i.e.} having large $T$ and $S$, possibly obviating the need for contrived cosmological boundary conditions. Evolution is dictated by the system's need to reduce its energy, which is achieved by lowering $T$ and expanding. This might define in parallel an ``arrow of time",  the increase in $S$   compensating  the entropy loss due to the formation of the spacetime ``crystal", possibly providing a hint towards an explanation of the 2nd law of thermodynamics.

There is a certain $T$ however for which the behaviour of the partition function $Z$ in equation \ref{partition} changes dramatically. This change proves to be crucial for our argument towards symmetry emergence. For low $T$ such that 
\begin{equation}
k_{B}T \leq  k_{B}T_{c} = J/\ln{2} \sim 1.4 J
\end{equation}

\noindent we find $p \geq 1/2$ and $w \geq1$, implying that large clusters consisting of many edges dominate over the smaller ones! Readers familiar with the q-state Potts model recognize in this expression $T_{c}$ given by $e^{\beta_{c}J} = (1+\sqrt{q})/\sqrt{q}$ for $q=1$. This  has a highly non-trivial and far-reaching impact on the topology of the network of nodes. Since for a network to have any sense we assume that $\min{(E_{i})} = c$, a  lattice like the one of $E_{8}$ with $c=240$ contributes much more to the partition function than the conventional ${\cal Z}^{8}$ lattice having $c=16$. At this point, it is important to realize that $c$ corresponds to the ``kissing number" discussed in the previous section.  Approaching $T_{c}$ therefore allows the possiblity of having  a $E_{8}$ lattice  emerge spontaneously from the dynamics.
The relevance of densest-sphere packing and ``highest kissing number" arguments presented in the previous section is now apparent, since this particular lattice offers an optimal configuration with regards to $c$ and might be preferred over alternative arrangements not offering so many edges per node.
 Note moreover that  ``crystal" clusters with $E_{i}=240$  evolve even when $T>T_{c}$. However, such clusters should lie within limited regions not contradicting Big-Bang nucleosynthesis.

At this critical point other lattices with even higher $d$ and  $c$ could also form. These however do not lead to the symmetries observed in our world, as explained in the last section. This implies that we might be living within a metastable region, with other Universe domains corresponding to different configurations of lattice points, devoid of the known interactions. This is consistent with Ostwald's rule in polymorphic and allotropic crystallography, according to which the least stable polymorphs crystallize first, leading to transformations between closest phases with regards to free energy. We return to this issue shortly.

Returning to the $E_{8}$ lattices now, near $T=T_c$ one finds a free energy $F_{C_{i}} \sim k_{B}(T-T_{c})\ln{\Omega_{C_{i}}}$ and an entropy $S_{C_{i}} \sim  - k_{B}\ln{\Omega_{C_{i}}}$, where $\Omega_{C_{i}} = 2^{2E_{i}}$ has a combinatorial interpretation, expressing the number of classically distinct configurations depending on whether a pair of nodes is connected or not by an edge. 
Since for  $T \equiv T_{c+}$ just above $T_{c}$ one expects $E_{i} \sim 1$, while for $T \equiv T_{c-}$ just below $T_{c}$ we have $E_{i} \sim 240$, the entropy $S(T)$ is discontinuous and the heat capacity $C \equiv T \partial S/\partial T$ is expected to diverge at $T=T_{c}$.  Moreover, at $T=T_{c}$ we expect a latent heat, or enthalpy, equal to 
\begin{equation}
H = T_{c}\bigl(S(T_{c+})-S(T_{c-})\bigr) = 478J.
\label{enthalpy}
\end{equation}
A more careful calculation of enthalpy might produce a smaller value for $H$ due to the action of the $Z_{2}$ discrete symmetries discussed in the previous section. 

Furthermore, another regime is $k_{B}T \ll J$, for which $F_{C_{i}} \sim  -E_{i}J  < 0  ~\rm{and}~  S_{C_{i}} \sim 0$. This describes the ground state and has a clear intuitive interpretation favouring large clusters. Zero entropy follows the 3rd law of thermodynamics. Negative energy implies positive pressure and contraction.  However, quantum corrections are here significant. The negative contribution stemming from cluster formation is nearly cancelled by the quantum mechanical ground-state energy, leading to roughly zero energy and pressure, and lattice points close to equilibrium.

Metastability of the two $E_{8}$ lattices is central to the argument presented and should be studied more thoroughly now. 
To proceed, recall that, in a sphere packing, the space $R^{d}$ can be geometrically divided into identical regions $F$ called fundamental cells, each of which contains just one sphere. Thus, the density of a sphere packing is given by
$\phi =v/Vol(F)$, where $v$ is the volume of a single d-dimensional sphere and $Vol(F)$ is the d-dimensional
volume of $F$ \cite{selforg}. The density $\phi$ is a decreasing function of $d$. Metastability is at least qualitatively supported from the fact that the $E_{8}$ lattices offer a local maximum for a properly-normalized kissing number (nKN) and for the centre density, defined as the ratio of sphere-packing density over the unit-sphere volume for certain dimensions $d$. Indeed, there exist  bounds according to which the $E_{8}$ lattice offers a centre density higher than the maximum achievable for $8<d<12$ \cite{centerdensity}. In parallel, regarding known sphere packings, it offers maximal centre density for $6<d<18$ and a local maximum for nKN \cite{kissingnu}. The next dimension  probably providing a local maximum for the centre density and for nKN hosts the even unimodular (i.e. self-dual) Leech lattice in $d=24$ offering maximal centre density for $0<d<28$ and a kissing number equal to 196,560. Although it exhibits symmetries not associated with the known Lie groups, (instead, it is closely connected to the largest sporadic finite simple group), one should explore further  the relevant Physics since it might correspond to a more stable equilibrium where our Universe might eventually decay into. 

To be more explicit, we assume that, near its ground state, 
 the behaviour of our system is dictated by a potential of the form 
${\cal V} =-\phi  E_{i}J$.  We define an optimal  potential ${\cal V}_{opt}$ as 
${\cal V}_{opt} \equiv -\phi_{max}E_{i~(max)}J$, where $\phi_{max}=2^{-0.6d} \geq \phi$ is a theoretical upper bound of $\phi$ for large $d$ \cite{selforg} and $E_{i~(max)}$ is the maximal lattice kissing number known in a given dimension. Even though the expression for $\phi_{max}$ gives a bound that is too stringent for lower values of $d$, using instead the largest known packing-density values for $d < 11$ does not spoil qualitatively the behaviour of the optimal potential.  The values of ${\cal V}_{opt}$ for $J=1$ in various dimensions are listed in Table 2. As can also be seen in the related Figure 2, ${\cal V}_{opt}$ exhibits two, almost degenerate, minima at $d=8$ and $d=24$. Since a known kissing-number configuration does not necessarily saturate the packing-density upper bound, the calculation of  ${\cal V}_{opt}$ above benefits suboptimal packings. This only makes our argument stronger, since the local extrema of the potential correspond to the $E_{8}$ and Leech lattices, which are most likely optimal.
This provides a first indication that our system is naturally lying in a metastable configuration dictated by the $d=8$ minimum. 
In order to prove the metastability of the $E_{8}$ lattice rigorously, one needs to improve further the upper bounds for kissing numbers in $R^{d}$ for $d>8$ \cite{Mittel}. Although present bounds suggest that $d=8$ is preferred over
 lower-dimensional configurations, presently-unknown configurations for $d>8$ (other than the Leech lattice) might correspond to kissing numbers rendering them energetically slightly more favorable.

\begin{figure}
\vspace{-4cm}
\includegraphics[width=1\textwidth]{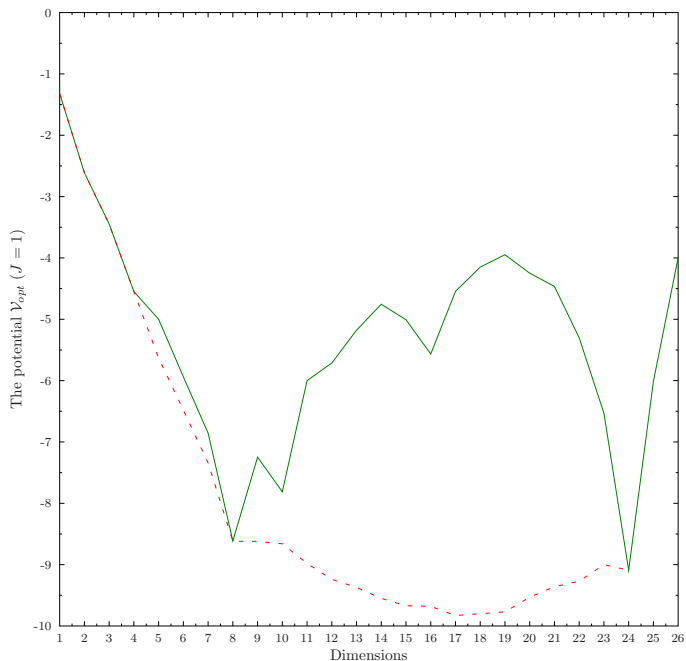}
\vspace{-4cm}
\caption{The potential ${\cal V}_{opt}$ ($J=1$) suggesting metastability of the $E_{8}$ lattice. The solid line corresponds to the known $E_{i~(max)}$ values listed in Table 1, while the dotted line corresponds to to the upper bound of $E_{i~(max)}$ given in \cite{Mittel}}
\label{fig:major} 
\end{figure}

\begin{table}[t]
\caption{$E_{i~(max)}$ and ${\cal V}_{opt}$ in various dimensions $d$. Kissing numbers 
for $8<d<16$ and $d>24$ correspond to non-lattice packings} 
\centering 
\begin{tabular}{lll} 
\hline\noalign{\smallskip} 
Dimension  $d$ & Kissing number $E_{i~(max)}$ & ${\cal V}_{opt}(J=1)$ \\ [0.5ex] 
\noalign{\smallskip}\hline\noalign{\smallskip}
1 & 2 ~~~~ ($SU(2)$ lattice $\sim {\bf Z}$)& -1.32\\ 
2 &  6 ~~~~ ($SU(3)$  lattice)& -2.61\\
3 &  12 ~~ ($SU(4) \sim SO(6)$ lattice)& -3.45\\
4 &   24 ~~ ($SO(8)$ lattice)&  -4.55\\
5 &   40 ~~ ($SO(10)$ lattice)& -5.00\\ 
6 &   72  ~~ ($E_{6}$ lattice)& -5.94\\
7 &   126 ~ ($E_{7}$ lattice)& -6.86\\
{\bf 8} &    240  ~ ($E_{8}$ lattice: optimal)  $<$even, self-dual$>$& -8.62\\
9 &   306 &  -7.25\\ 
10 &   500 &  -7.81\\
11 &   582 &  -6.00\\
12 &   840 & -5.71\\
13 &   1154 &  -5.18\\ 
14 &   1606 & -4.75\\
15 &   2564 & -5.01\\
{\bf 16} &   4320 (Barnes-Wall lattice) $<$even, not self-dual$>$&  -5.57\\
17 &   5346 &  -4.54\\ 
18 &   7398 &  -4.15\\
19 &   10668 &  -3.95\\
20 &   17400 &   -4.25\\
21 &   27720 & -4.46\\ 
22 &   49896 &  -5.30\\
23 &   93150 &  -6.53\\
{\bf 24} &  196560~ (Leech lattice: optimal) $<$even, self-dual$>$&  -9.09\\
25 &   197040 &   -6.01\\ 
26 &   198480 &  -4.00\\ 
\noalign{\smallskip}\hline 
\end{tabular}
\label{table:major} 
\end{table}

A more detailed study of critical behaviour in this context clearly necessitates computer 
simulations, which have proven to be indispensable even for much simpler physical systems. In particular, it would be very useful to further study the metastability of the various possible lattices, in order to check the validity of the emergence scenario presented. For
 temperatures close to criticality, it might prove to be useful to also study the behaviour of the 
$q=2$ Potts model as being closer to the fermionic action we started with. Moreover, one should study further the mechanism by which the two $E_{8}$ lattices couple to each other during the ``crystallization" and compactification process, in a way consistent with the discussion of the previous section. We close this section by noting that, in a space that is inherently discrete as the one just analysed, the continuum limit required in order to study consistently the $E_{8}$-singlet fermion condensate is ill-defined. Therefore, the antisymmetric $\textbf{248}_{a}$ fermion condensate considered in the previous section is left as the most natural trigger of symmetry breaking.

\subsection{Cosmological implications}
 Next, we describe some potentially interesting cosmological  implications of this critical behaviour. We take the measured  $\Lambda \sim 2 \times 10^{-3}$ eV to correspond to the free energy of a minimal cluster of the $E_{8}\times E_{8}^{\prime}$ lattice having 240 edges from each of the two $E_{8}$'s, {\it i.e.} $E_{i}=480$, and  $T_{CBR} \sim 2.7 \times 10^{-4}$ eV of the cosmic background radiation to be equal to the system's temperature. Then,  the critical free-energy expression  yields
$\Lambda \sim   7 k_{B}T_{CBR} \sim  10J$ and 
$\epsilon \equiv  (T_{CBR} - T_{c})/T_{c} \sim 1\%$. This implies a certain fine-tuning close to criticality for $T$. Taking the Universe to be in a ``glass-to-crystal" transition, typical relaxation and equilibration times $\tau$ for glassy dynamics are huge compared to the microscopic ones of ferromagnetic-type systems. This leads to considering non-adiabatic phenomena, since $\tau$ is given by $\tau \sim \xi^{z} \sim (T-T_{c})^{-\nu z}$, with $\xi$ the correlation length and $\nu, z$ critical exponents. Near the transition point, $\xi$ and $\tau$ diverge. This implies that relaxation times are of cosmological scale, and we might be  just living within such a critical period.

Another issue to address is the smallness of $\Lambda$ in comparison to $M_{\rm{Planck}}$. The solution to this puzzle might be coming from the Hamiltonian we started with, which includes only local, nearest-neighbour, interactions, effectively introducing a very large infrared cut-off.  For this interpretation to work,  Feynman integrals are to be performed over the whole momentum space only when particles are present in the Feynman diagram, which can exhibit non-local behaviour. Spacetime itself is local and momentum integration should not be allowed to reach values much lower than $M_{\rm{Planck}}$.
Alternatively, one can introduce a phenomenological potential $V(r)$ between lattice sites being separated from each other by a distance $r$, similar to the Lennard-Jones type, given by
\begin{equation}
V(r) = -2E_{i}J\left(\frac{L_{\rm{Planck}}}{r}\right)^{d-3}\Biggr(1-\frac{\sqrt{Ja}}{2}\left(\frac{a}{r}\right)^{d}\Biggr)
\end{equation}

\noindent with $L_{\rm{Planck}} = 1/M_{\rm{Planck}}$, $d=16$ and $a$ having dimensions of length corresponding to the distance where ``repulsive" effects become important. For large $r$, $V(r)$ vanishes like $1/r^{d-3}$, since the 2-point correlation function in position space falls as $ 1/r^{d-2}$. Using the fact that $d \gg 1$,  min$(V(r)) \sim -E_{i}J$  in accordance with the ground-state free energy reached for 
$r_{min} \equiv L_{\rm{Planck}} \sim a (Ja)^{1/2d}$
defining in parallel  $L_{\rm{Planck}}$ as a function of $J$. 
 The potential exhibits a non-linear behaviour with respect to $J$, a situation possibly traceable back to the  action of equation \ref{action} before linearization. Moreover, $V(r)$  has a zero at $r = 2^{-1/d}L_{\rm{Planck}} \sim L_{\rm{Planck}}$, {i.e} close to the value where it has its minimum. 

Since the slope of $V(r)$ for small $r$ is proportional to a large power of $a$, adjusting the modulus $a$ can control  the steepness of the potential. 
 Using the relation 
 \begin{equation}
JL_{\rm{Planck}}=\left(\frac{L_{\rm{Planck}}}{a}\right)^{2d+1}
\label{cosmohier}
\end{equation}

\noindent derived directly from the minimization condition and  taking $a \sim 9.3 L_{\rm{Planck}}$  gives the required hierarchy $\Lambda \sim 10^{-31}$ $M_{\rm{Planck}}$ between $\Lambda$ and $M_{\rm{Planck}}$. The extent of this hierarchy might therefore be traced not only  in the large dimensionality of our space,  but also in the steepness of the repulsive potential between sites.
Moreover, the quantity $JL_{\rm{Planck}}$, apart from falling with increasing $a$, is  proportional to the system's Vol$_{c}$ and inversely proportional to the emergent gauge coupling. Symmetry breaking could then be associated with a shrinking Vol$_{c}$ and with a critical value of the steepness parameter $a$, {\it i.e.} $a_c \sim 9.3 L_{\rm{Planck}}$, above which fermion condensates form. Anyway, the form of the 2nd term of the potential  needs  further justification, probably in terms of a series expansion in powers of $(a/r)^{d}$, where the value of $a$ might be determined by the geometry of the lattice.

 Consequently, a higher-dimensional analogue of spin-glass phase transitions might provide a picture for the emergence of $E^{}_8 \times E_{8}^{\prime}$ at the beginning of our Universe, as a kind of ``liquid-to-solid", freezing phase transition, or a  kind of disorder-order, ``glass-to-crystal" transition. Regarding entropy $S$, the only way for the system to compensate for the loss of $S$ within a spacetime volume $Vol$ during a time $dt$ is to expand, changing its volume by $d(Vol)$. This leads to an equation $Vol = (Vol)_{0} \exp{(\mu E_{i}t)}$ with $\mu$ constant. Although large values for $E_{i}$ imply a period resembling inflation, a study in this direction exceeds the bounds of the present analysis. Other cosmological implications include the existence of macroscopic domains in the Universe not having the symmetries  observed in our neighbourhood. Particles within such regions would not interact in familiar ways, for instance not feeling electromagnetic interactions and possibly supplying an explanation for Dark Matter  (DM). The luminous parts of galaxies would occupy regions corresponding to the ``jammed", ordered  phase of spacetime, domain states of ferromagnetic type, like ``crystal bubbles" within a glass-type, amorphous spacetime structure. Domain growth would  be described by a relation of the form $\xi(\tau) \sim  \tau^{1/z}$. The ratio $R$ of crystal-to-glass-type volumes would be given by $R = 1 - \exp(-\Delta F/k_{B}T)$, where $\Delta F$ is the free energy gained by the system by being in the ``crystal" state. Nucleation and growth of crystalline grains within amorphous glass materials is a frequently-studied subject in solid-state physics \cite{glass} and could  provide a testing ground for related cosmologies. 

A related scenario that could be analysed might predict that ``spacetime" nucleation continues today, implying a growth of the luminous-to-DM ratio on cosmological time-scales. Using the expression for the critical free-energy, we find $R = 1 - 2^{-2\epsilon(480-\tilde{E_i})}$ where $\tilde{E_i}$ is the number of edges, per potential $E^{}_{8}\times E_{8}^{\prime}$ lattice cluster, of the ``glass" state. For $\epsilon \sim 1\%$, we find the following possible characteristic $(R, \tilde{E_i})$ pairs: $(5\%, 476)$, $(24\%, 460)$ and $(75\%, 380)$. A detailed analysis towards this direction would allow the prediction of galactic DM concentrations and of the average structure of the underlying spacetime lattice. It would allow answering questions like ``is the structure of spacetime within the intergalactic voids of glass- or crystal-type?"  and compare results with DM considerations \cite{dark}. 
Assuming that ``crystal" domains are occupied by visible galaxies implies that $R \sim 5\%$, while taking intergalactic voids to be also ``crystal"-like raises $R$ to around $77\%$. Other possibilities include DM regions corresponding either to alternative $E_{7}$ symmetry breakings, or to denser ``sphere packings" not linked to a group's root system. The latter takes us to a scenario where our vacuum has already started decaying towards a more stable configuration like the Leech lattice, leading to growing DM and shrinking luminous domains.

Alternatively, in off-equilibrium phenomena of crystal and glass formation, the fluctuation-dissipation theorem in fast transitions is violated, since the system does not have enough time to relax to its new equilibrium, forcing us to consider ``effective temperatures" $T_{eff}$ even an order of magnitude larger from the heat-bath ones \cite{relaxation}. Here, $R \sim 50\%$  implies $T_{eff} \sim 10T_{CBR}$, while $R \sim 80\%$ implies $T_{eff} \sim 5T_{CBR}$. If the universe expanded and cooled too fast to have relaxed to equilibrium, $T_{eff}$ for ``crystal" formation is larger than $T_{CBR}$. This might  explain expansion or inflation in terms of $\Delta F = \Delta F(t)$. A disordered initial configuration in a  ``liquid" state has a  higher energetic gain by forming a crystal than a ``glass" state. As the Universe cools down and ordered ``glass-type" structures emerge, $\Delta F(t)$ decreases. Such values of $T_{eff}$ are consistent with treating particles as topological or vacancy defects on the lattice background described above, analogous to positively-charged holes in an electron sea or lattice, their number density $d_{p}$ in the Universe being a function of enthalpy cost and approximately equal to the ratio of their entropy $S_{p} \sim k_{B}10^{89}$ to the Universe entropy $S_{B} \sim k_{B}10^{122}$ assuming a Bekenstein-bound saturation \cite{Astro}, {\i.e.} 
\begin{equation}
d_{p} \sim S_{p}/S_{B} \sim 10^{-33} \sim \exp{(-H/k_{B}T_{eff})},
\end{equation}
\noindent  implying from equation \ref{enthalpy} that $T_{eff} \sim 4.4 T_{CBR}$. This result favours in parallel the characterization of intergalactic voids as ``crystal-like". Note that the discrepancy between $T_{CBR}$ and $T_{eff}$ might actually be smaller if a more careful calculation of the enthalpy $H$ is performed.

Moreover, the position of galaxies and such ``crystal bubbles" might be correlated, with mass acting as a topological defect closely connected to the formation of a spacetime ``crystal". This has far-reaching implications on the structure formation of galaxies,  consistent with the view that stars are born within DM halos.  It could potentially lead to an understanding of the shape of spiral galaxies on the basis of ``helicoidal dislocations" in crystals. It might also explain the large voids between galactic clusters, since crystals usually displace impurities towards boundaries of different phases and form vacancy clusters  to minimize their energy. In our case, the role of 
impurities is played by DM regions containing small crystal ``islands", {\it i.e.} galaxies. The latter are full of vacancy defects which are ``frozen-in" during  cosmological expansion, in analogy to a similar phenomenon occuring during fast crystallization, and which correspond to elementary particles. Such considerations might also solve the ``dwarf galaxy" problem, {\i.e.}  the rarity of ``dwarf" galaxies, which are an order of magnitude less than predicted by simulations \cite{dwarf1}\cite{dwarf2}, since in these ``dwarf" galaxies, particle density, seen as a defect, has not reached values consistent with nucleation and ``crystal" formation. Experiments could be designed  in the far future to probe the spacetime structure within DM domains, or to measure the potential energy release, perhaps in the form of ultra-high energy cosmic rays, which are presently of unknown origin, when the ``crystal" forms. However, critical phenomena of this kind are intractable even in low dimensions, necessitating the use of phenomenological potentials and simulations in the area of glass-to-crystal transitions which is still far from being well understood. Although a thorough analysis in this direction transcends our purposes, the smallness of the luminous-matter  portion of the total energy of the Universe makes us reluctant to discard so radical solutions of the DM and Dark-Energy puzzle too hastily. We close this discussion by noting a similar attempt to circumvent problems cold-DM models have with galactic mass distribution by positing that DM is topological \cite{Kirillov}. 

Next, we discuss briefly quantization of our action. What follows is  speculative, noting a proximity of the present theory  with some current approaches to quantum gravity. 
The fact that the group $E_8$ is simply laced allows the definition of a common scale 
$1/M_{\rm{Planck}}$ equal to the roots norm, an identification defining a fundamental scale for the 
theory. 
By having $\Gamma^{}_8 \times \Gamma_8^{\prime}$ emerge with lattice spacing equal to $L_{\rm{Planck}}$, one  achieves a cellular decomposition of spacetime with a UV cut-off  equal to $M_{\rm{Planck}}$, avoiding in principle singularities plaguing quantum gravity.  The number of faces, edges and vertices of the various simplices  is determined by this lattice. This might have a dramatic impact not only on the general renormalization programme but also on black holes, the initial singularity of spacetime and gravitational collapse,  analogous for some to the false prediction of atom collapse before the advent of quantum mechanics.

Moreover, the metric in section 2  is reminiscent of the one in the spinorial version of Ashtekar variables \cite{Ashtekar} using no background metric. In an approach close to spin networks and lattice Yang-Mills, we have lattice nodes corresponding to 4d spacetime points, a ``world crystal", where a  continuum perturbative limit is ill-defined since spacetime is inherently discrete, while flat Minkowski space lies far from perturbative considerations. On each node there is a fiber corresponding to $E^{}_7 \times E_{7}^{\prime}$ stemming from 14 compactified dimensions.
A  probability wavefunction extending between 2 adjacent lattice nodes corresponds to a particle with Planck-scale energy, which seems a reasonable configuration for the first moments after the creation of the Universe. It also provides an understanding of Heisenberg's uncertainty principle, since a particle is not localized on a node but  extends at least between 2 adjacent nodes. 
 This is compatible with treating particles with spin as spacetime  defects like ``dislocations" and ``disclinations" in  Einstein-Cartan action theories \cite{Petti}\cite{Kleinert}, reminds us of Wheeler considering particles as ``quantum geometrodynamic excitons" \cite{Preparata} in an analogy inspired by solid-state Physics
 and is consistent with our previous treatment of particles as topological or vacancy defects. 
 Such schemes might also lead to a small $\Lambda$ due to a kind of ``equilibrium" between lattice sites \cite{Kleinert}. 
The above provide just a heuristic hint towards quantization, more work being required  to derive  robust results.
Experiments around $M_{\rm{Planck}}$ should distinguish such a spacetime fabric from models treating  particles as extended objects on a continuous spacetime background. Ideas along these lines might include in the future a gravitational analogue of Bragg spectroscopy  probing the microstructure of spacetime. 

\section{Discussion}
\label{sec:6}
\subsection{Open issues}
Considering the results of the previous sections, we have to note that several open issues still remain. Most important, the issue of reproducing reliably the equations of General Relativity starting from an action of the form given in equation \ref{action} and the issue of providing a solid framework leading to its quantization, while the emergence of a UV cutoff in the theory is an encouraging sign in this direction. Next, one 
should  show that the 16d torus over the $\pi\Gamma_{8}\times \pi\Gamma_{8}$ lattice emerges 
naturally and produces the action needed to describe correctly Physics at long wavelengths, 
including the number of internal dimensions, their possible relation to inflation, and the value of the cosmological constant. In this respect, the metastability of this lattice should be studied further, since it would offer a solid argument for the emergence of the symmetries observed in nature. In addition, more work is  required towards proving that the Potts lattice model analyzed belongs to the same universality class as the one required by our spinor-gravity approach, in order to draw safer conclusions on the validity of the qualitative cosmological  implications of this model, including the interpretation of elementary particles as vacancy defects, or ``defectons" in a quantum setting. Then, a  justification of the
 parity-violating assumption $g_F >> g_K$  is needed, probably in relation with the  
compactification volumes. Furthermore, a rigorous justification of the breaking channels of the 
symmetries down to the SM should be sought, since these channels, although attractive, are not unique. In addition, a more reliable calculation of the critical coupling associated to the breaking of the initial symmetry should be performed. 
Moreover, unitarity issues have to be tackled because of the non-compact form of the Lorentz 
group $SO(3,1)$.
In principle, a metric-independent topological action like the one in equation \ref{action} should cure 
such issues when considered from the viewpoint of a fundamental theory. 

On the experimental front, a problem with the fermion content described in this work is that it is usually associated with an $S$ parameter which is larger than what is measured experimentally. It remains to be seen whether non-perturbative vertex corrections in such models  can drive the $S$ parameter close to zero, although so many new fermions  are introduced \cite{Triantap3}. On the other hand, some LEP and Fermilab results at the $2-3 \sigma$ level are in principle compatible with  quark-katoptron mixing and the existence of katoptron bound states \cite{Triantap5}, since they suggest mass-dependent anomalous quark couplings \cite{Renton} and an excess of dijet plus W-boson events \cite{recentFermi}. Moreover, assuming that the theory exposed above is correct, recent LHC results regarding a new boson having a mass of around 126 GeV correspond to the lightest member of a series of katoptronic mesons, similar to QCD or technicolor mesons. This lightest meson, a ``katoptronic pion" corresponding to the lightest katoptrons, is expected to have comparatively very small 
couplings to third-generation ordinary fermions, like the bottom and top quark, due to the comparatively small mixing of the lightest katoptrons to third-generation ordinary fermions. We expect future collider experiments in LHC and elsewhere  to shed more light on some of these issues by carefully studying the fermionic couplings of any new bosons detected.  

\subsection{Conclusions}
We exposed above an attempt to unify gauge with gravitational interactions using  $E^{}_{8}\times E_{8}^{\prime}$ emerging naturally from first principles. It presents several advantages, not requiring many arbitrary parameters, nor fundamental scalar fields, nor extra space dimensions; it leads to coupling unification and to an understanding of the unification coupling strength from an invariant of the emergent symmetry group; it reproduces the symmetries, the family structure of matter and the dimensionality of spacetime, far from treating it as background; it provides a possible solution to the hierarchy problem between the Planck scale, the weak scale and the cosmological constant scale. Moreover, it exhibits a unique vacua sequence with cosmological implications like the interpretation of DM as having a topological origin. Securing the present approach on a firm basis needs, among many other things, a new physical principle which refers to ``optimal connectivity". This principle lies presumably at the heart of several other scientific areas as well (like crystallography, cognitive science etc.) and is more fundamental than a given spacetime or gauge symmetry; according to it, spacetime, matter, and their symmetries, emerge naturally, after a relevant phase transition, from a set of identical, distinct elementary fields connected to each other optimally. 

 
\end{document}